\documentstyle[aps,epsf,twocolumn]{revtex}
\tighten
\begin{document}
\draft
\twocolumn[\hsize\textwidth\columnwidth\hsize\csname
@twocolumnfalse\endcsname


\title{Late time dynamics of scalar perturbations outside black holes.\\
                         I. A shell toy-model}
\author{Leor Barack\cite{Email}}
\address {Department of Physics,
          Technion---Israel Institute of Technology, Haifa, 32000, Israel}
\date{\today}
\maketitle


\begin{abstract}
We present a new analytic approach for the study of late time evolution of
linear test-fields, propagating on the exterior of black holes.
This method provides a calculation scheme applicable to Kerr black holes
(for which case no analytic calculation of the late time tails has
been presented so far).
In this paper we develop the new technique and apply it to the case of
massless scalar waves evolving on the background geometry of a static
spherically-symmetric thin shell with a Schwarzschild exterior.
The late time behavior of the scalar field at null infinity is calculated,
and is explicitly related to the form of  (quite arbitrary) initial data.
This reproduces the well-known late time power-law decaying tails.
In an accompanying paper we apply our approach to the complete Schwarzschild
black hole geometry, where we obtain the familiar inverse-power late time
tails  at null infinity, as well as at time-like infinity and along the
event horizon.
A detailed calculation of the late time power-law tails in the Kerr geometry,
based on the same approach, will be presented in a forthcoming paper.
\end{abstract}

\vspace{6ex}
]

\section{introduction} \label{secI}

The gravitational field outside black holes created during a generic
gravitational collapse relaxes to the stationary
Kerr-Newman field. This result, referred
to as the ``{\em no-hair theorem}'' (see, for example, \cite{Heusler96})
implies that outgoing radiation should carry away all the initial
characteristics of the collapsing object, except for its mass, charge, and
angular momentum. These three quantities, characterizing the Kerr-Newman
field, are conserved by conservation laws; any other quantity will
vanish by the time the gravitation field settles down on its
stationary state.

The underlying mechanism for this relaxation process was first demonstrated
by Price \cite{Price72} for the case of a nearly spherical collapse. Price
analyzed the dynamics of massless integer-spin test fields, evolving on the
background of a Schwarzschild black hole (SBH), and showed
that when viewed from a fixed location outside the black hole, the
waves die off at late time with an inverse power-law tail (in the
Schwarzschild time $t$), whose power index depends only on the
multipole
number $l$ of the mode under consideration. Later \cite{Gundlach94I}, it was
demonstrated by
analytic and numerical methods that the nearly spherical
collapse exhibit late time decay tails also at future null infinity and
along the event horizon. The formation of these tails was explained
as due to back-scattering of the outgoing radiation off spacetime curvature
at very large distances.

The existence of late time tails was demonstrated for linear perturbations
of both SBH \cite{Price72,Gundlach94I} and
Reissner--Nordstr\"{o}m \cite{Gundlach94I} exteriors.
Remarkably, numerical analysis of the fully non-linear dynamics of
fields yields the same decay rates at late time, as for the
minimally-coupled (linear) fields \cite{Gundlach94II,Burko97}.
This, of course, encourages the application of the perturbative approach,
even though the problem in discussion is non-linear in its nature.

Ching {\em et al.\ }\cite{Ching95} explored a wide class of
asymptotically-flat
spherically symmetric spacetimes, represented by effective curvature
potentials of the form $(\ln r)^{\beta}/r^{\alpha}$ (where $\beta=0,1$ and
$\alpha>2$ are parameters). This class includes the SBH and
Reissner-Nordstr\"{o}m geometries.
They found that, generically, the wave behavior at late time is
not characterized by a strict power-law tail, but rather it has the
form $(\ln t)^{\beta}\times$(inverse-power in $t$).
In that respect, the Schwarzschild and Reissner-Nordstr\"{o}m geometries
represent a special subgroup of spacetimes.
For the monopole moment of the scalar radiation,
it was shown by G\'{o}mez {\em et al.\ }\cite{Winicour94} that
the form of the tails (whether "logarithmic" or not) depends on whether
or not the Newman--Penrose constant for the field vanishes.

Recently \cite{Brady97}, Brady {\em et al.\ }studied scalar waves dynamics
in the non-asymptotically-flat exteriors of Schwarzschild--de Sitter
and Reissner--Nordstr\"{o}m--de Sitter black holes. Contrary to the
asymptotically-flat geometries, no power-law tails were detected in these
cases. Instead, the waves were found to decay exponentially at late time.

The aforementioned analysis by Ching {\em et al.\ }follows a technical
scheme, first
introduced by Leaver \cite{Leaver86}, in which the linear waves are first
Fourier-decomposed, then evaluated in the complex frequency plane
(see also \cite{Andersson97}).
In this technique, the late-time tails are explained in terms of a branch
cut in the Green's function in the frequency domain.
The fact that the branch cut is due to the form of the potential at
asymptotically large radius (see \cite {Ching95} for details) implies,
again, that the tails originate from scattering off the curvature potential
at large distances. This observation, in turn, suggests that the development
of tails is independent of the existence of an event horizon. Thus
the tail phenomenon appears to be of a more universal nature: it may
characterize the realistic stellar dynamics. This, indeed, was suggested
by Gundlach {\em et al.\ }\cite{Gundlach94I,Gundlach94II} and Ching
{\em et al.\ }\cite {Ching95}.
In \cite{Gundlach94II}, for example, the
purely spherical collapse of a self-gravitating minimally coupled scalar
field was studied. It was demonstrated numerically that in this case late
time
tails are formed even when the collapse fails to create a black hole.
We shall further discuss this issue in the present paper.

Historically, the study of wave dynamics outside black holes was
motivated by the will to construct a detailed description of the relaxation
process leading to the stationary ``no hair'' state (For example, there was
a need to verify that the event horizon is indeed stable under generic
perturbations). Many recent studies, however, are oriented by the prospects
of directly observing gravitational radiation from astrophysical systems
(e.g. with the LIGO observatory, see \cite{Thorn87}).
Although the study of late time behavior of fields outside black holes has
probably no immediate observational implications (late time outflux
from a realistic coalescing binary system, for example, would be orders of
magnitudes weaker than the short pulse of radiation expected during the
last few seconds of the merger), it still addresses several important
questions. For example, this study has crucial relevance to the exploration
of the internal structure of black holes, as it provides the input for the
internal wave evolution problem. Particularly, it has been argued
\cite{Ori91} that the form of the slowly decaying wave tail along the event
horizon affects the strength of the mass-inflation singularity at the Cauchy
horizon inside charged and rotating black holes.

The study of waves evolving in curved geometries is also interesting on its
own right, from a theoretical point of view: such waves,
massless just as well as massive, do not propagate along light cones solely;
rather they also spread inside them. Regardless of the presence or absence
of
an event horizon, it is this feature of the evolution which is responsible
for the phenomenon of late time decay tails in curved spacetimes.

In virtue of previous studies, we now have the following schematic picture
regarding the dynamics of
waves outside a {\rm nearly-spherical} collapsing object: Consider a
perturbation in the form of a compact pulse of radiation (gravitational
or electromagnetic), somewhere
outside the collapsing object, emitted at some time during the collapse.
This
pulse may represent radiation emerging from the surface of a collapsing
object, as well as any other form of perturbation on the background geometry.
We shall refer to it as the ``initial pulse''.
A static observer outside the black hole will then indicate three successive
stages of the wave evolution.
First, the exact shape of the waves front depends on the detailed form of the
initial pulse. This stage (lasting a period of time comparable to the
duration of the initial pulse) is followed by a ``quasinormal ringing''
(QNR) relaxation stage, during which the
waves undergo exponentially-decaying oscillations with (complex)
frequencies completely determined by the mass and electric charge of the
central object \cite{Chandra83}. Finally, as the QNR
dies off exponentially in time, it leaves behind an inverse
power-law decaying tail of radiation. During the last two stages, the details
of the initial pulse affect the shape of the waves only through a global
amplitude factor, hence the evolution during these stages is purely
characteristic of the background geometry.

Now, a realistic black hole formed by a generic gravitational collapse is
expected to spin, as do astrophysical stars (On the contrary, models of black
holes with electric charge and/or cosmological constant are mainly
hypothetical). Therefore
it is natural to ask how the tail phenomenon is affected by the presence of
angular momentum in the background geometry. One may suspect that the late
time behavior of waves,
evolving outside a rotating Kerr black hole, would be qualitatively
of the same nature as in the spherically-symmetric spacetimes. This is
because the Kerr geometry asymptotically approaches that of a
SBH at very large radius, and presumably it is this far region of spacetime
whose structure determines the form of the late time radiation.

Yet, in spite of the obvious interest, no analytic scheme has been
proposed thus far to describe wave dynamics outside rotating black holes.
The basic obstacle is, of course, the fact that the
axially-symmetric Kerr black hole possesses three non-trivial
dimensions, instead of only two in the spherical cases. This makes both
analytic and numerical investigation significantly more complicated.
Recently, there was an initial progress, with the introduction of a full
($2+1$ dimensions) numerical analysis by Krivan {\em et al.\ }
\cite{Krivan96,Krivan97}, though much effort is still required is this
direction.

\subsection*{The new calculation approach}

We have developed an analytic calculation scheme enabling one to analyze late
time wave dynamics outside Kerr black holes.
This analysis shall be presented in \cite{Ours} (see also
\cite{Barack97}).
The major goal of this paper,
together with the accompanying paper (to be referred to as {\bf Paper II}),
is to demonstrate the applicability of our technique in a simpler model,
namely the evolution of scalar waves in the SBH exterior. This will serve
several purposes: First, we shall be able to test our scheme against the firm
results already obtained for this case by previous studies. Secondly, many
parts of the formalism to be developed shall be later directly employed in
the analysis in Kerr.
Finally, our analysis in Schwarzschild proves to be valuable on its own
right, providing, in some respects, a more complete picture of the late-time
wave behavior than already available.

Basically, our analysis is composed of two major steps. In the first and more
crucial one, a characteristic initial value evolution problem for the scalar
waves is treated analytically, resulting in the construction of a late-time
solution for the wave {\em at null infinity}.
This calculation involves the introduction of a special perturbative
decomposition of the waves, followed by the application of
the standard (time domain) Green's function technique.
In the second step we then use a simple
late time expansion of the wave near time-like infinity, in
order to obtain the late time behavior of the wave at {\em any} constant
radius,
including along the event horizon. This second step is made possible only
after the waves' form at null-infinity is derived by the first step of the
analysis.

To introduce the analysis at null infinity (i.~e.\ the first step mentioned
above) in a clear and more instructive way, we incorporate in this paper a
simple
toy-model, in which the scalar waves are taken to evolve in the
gravitational field
induced by a spherically-symmetric thin shell of matter. This configuration
possesses the same spacetime structure at large distances (outside
the shell) as in a complete SBH manifold of the same mass, while
its small-$r$ structure is much simpler.
That would reduce the amount of technical details to deal with when
developing
our calculation scheme, and may also enable one to push the analytic
calculation
to a further extent, while leaving the essential features of the analysis
unaffected.
Indeed, we later show in {\bf paper II}, considering the complete SBH
manifold,
that the asymptotic late time behavior calculated at null-infinity is
unaffected
by the structure of spacetime at small radii, thus it is essentially the
same in
the complete SBH manifold as it is in the shell toy-model.

\subsection*{Arrangement of this paper}

In this paper we study the thin-shell toy model. To that end we
introduce our calculation method, which we call ``{\em the iterative
scheme}'', and apply it to calculate the behavior of the wave at
null-infinity in this model.

The paper is arranged as follows: In sec.\ \ref{secII} we
formulate the problem of wave evolution in the shell model as a
characteristic
initial-value problem. The iterative scheme is presented in sec.\
\ref{secIII}. It is applied in sections \ref{secIV} through
\ref{secVIII}, revealing the power-law pattern of the late time
wave decay at null infinity. The amplitude of the wave at late time
is explicitly calculated, expressed in terms of a (quite arbitrary)
initial data function.

\section{Initial-value formulation of the wave
         evolution problem}                      \label{secII}

\subsection{The shell model}\label{subsecIIA}
We consider a static spherically-symmetric thin shell of matter,
of a mass $M$ and some radius $R>2M$. The parameter $R$ should be
regarded as being of order $\sim 2M$ (say, $R=3M$). For this
configuration, the exterior vacuum region $r>R$ is a part of the
Schwarzschild spacetime, described by the line element
\begin{equation} \label{eq7}
ds^{2}=-f(r)dt^{2}+f^{-1}(r)dr^{2}+r^{2}(d\theta^{2}+\sin^{2}\theta
d\varphi^{2}),
\end{equation}
where $t$, $r$, $\theta$ and $\varphi$ are the standard Schwarzschild
coordinates and $f(r)\equiv (1-2M/r)$.
The region inside the shell, $r<R$, is flat:
\begin{equation} \label{eq8}
ds^{2}=-dT^{2}+dr^{2}+r^{2}(d\theta^{2}+\sin^{2}\theta d\varphi^{2}).
\end{equation}
Expressed in these coordinates, the geometry suffers a discontinuity at
$r=R$ (the metric functions $g_{rr}$ and $g_{tt}$ jump through this surface.
Also the time coordinates do not agree on both sides of the shell).

To allow a continuous representation of spacetime geometry, we define the
coordinates
\begin{equation} \label{eq9}
r_{*}=\left\{\begin{array}{ll}
          f(R)^{-1/2}\,r                                        & , r<R\\
          r-R+Rf(R)^{-1/2}+2M\ln\left(\frac{r-2M}{R-2M}\right)  & , r>R
        \end{array}
      \right.
\end{equation}
and
\begin{equation} \label{eq10}
t_{*}=\left\{\begin{array}{ll}
             f(R)^{-1/2}\,T   &  , r<R\\
             t                &  , r>R .
        \end{array}
      \right.
\end{equation}
In terms of the coordinates $t_{*}$, $r_{*}$, $\theta$ and $\phi$ the
geometry is described as a single continuous manifold.
(These coordinates form an example
of a `natural' set of coordinates through the `surface layer' $r=R$, as
discussed by Israel in \cite{Israel66}. The jump in the derivatives of the
metric functions through the shell is then related to the surface energy
tensor of this layer.)
Note that outside the shell, the radial coordinate $r_{*}$ is the
usual Schwarzschild's `tortoise' coordinate, satisfying
$dr_{*}/dr=f^{-1}(r)$.

\subsection{The initial-value problem}\label{subsecIIB}
We consider the evolution of initial data, representing a generic pulse of
massless scalar radiation, on the fixed background of the shell
described above.
The scalar field is assumed to satisfy the (minimally-coupled) Klein--Gordon
equation\footnote
{We comment, however, that other wave equations are also possible.
For example, to assure conformal invariance, one should rather use the
equation $\Box\Phi+\frac{1}{6}R\Phi=0$, where $R$ is the Ricci scalar. This
equation reduces to Eq.\ (\ref{eq1}) when considering waves in vacuum.}
\begin{equation} \label{eq1}
\Phi_{;\mu}^{\ ;\mu}=0,
\end{equation}
where $\Phi$ represents the scalar wave.
The structure of spacetime affects the evolution of the scalar wave through
the covariant derivatives, denoted in Eq.\ (\ref{eq1}) by semicolons.

Decomposing the field in spherical harmonics,
\begin{equation} \label{eq2}
\Phi(t,r,\theta,\varphi)=\sum_{l=0}^{\infty}\sum_{m=-l}^{l}\phi^{l}(t,r)
Y_{lm}(\theta,\varphi),
\end{equation}
and substituting in the wave equation (\ref{eq1}), we obtain an independent
equation for each of the components $\phi^{l}(t,r)$.
(We use the superscript $l$ to denote the multipole
number of the mode under consideration.)

A convenient form for the wave equation may be obtained in terms of a
new wave function \mbox{$\Psi^{l}(t,r)\equiv r\phi^{l}(t,r)$}.
To that end we
introduce the double-null (Eddington--Finkelstein) coordinates
$v\equiv t+r_{*}$ and $u\equiv t-r_{*}$.
The equation governing the evolution of the $l$-mode of the scalar wave
then reads
\begin{equation} \label{eq4}
\Psi_{,uv}^{l}+V^{l}(r)\Psi^{l}=0,
\end{equation}
in which the function
\begin{equation} \label{eq5}
V^{l}(r)\equiv\left\{\begin{array}{ll}
                \frac{1}{4}l(l+1)r_{*}^{-2}                     & , r<R\\
                \frac{1}{4}\left(1-\frac{2M}{r}\right)
                \left[\frac{l(l+1)}{r^{2}}+\frac{2M}{r^3}\right]
                & , r>R
            \end{array}\right.
\end{equation}
serves as an effective potential for the scalar field.
The form of the potential function completely determines
the effect of spacetime curvature on the wave evolution [terms proportional
to $M$], as well as the
centrifugal effect on the non-spherical ($l>0$) modes of the wave [terms
proportional to $l(l+1)$].

\begin{figure}[htb]
\input{epsf}
\centerline{\epsfysize 6cm \epsfbox{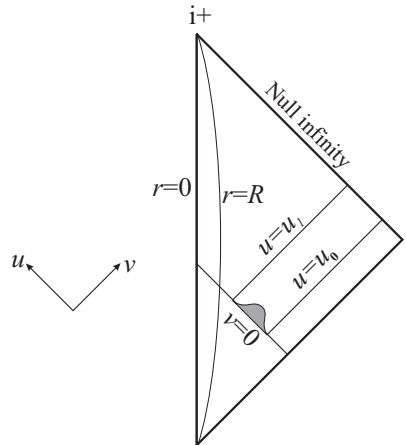}}
\caption{\protect\footnotesize The set-up of initial data.
         Shown is the conformal diagram representing spacetime
         geometry in the shell model.
         The spacetime region outside the shell $(r>R)$ is a part of the
         Schwarzschild
         manifold, while the interior of the shell ($r<R$) is flat.
         $u=u_{0}$ and $v=0$ are two initial characteristic (null)
         surfaces.
         An initially compact outgoing pulse is specified on $v=0$.}
\label{fig1}
\end{figure}

To set-up the initial value problem for the scalar
radiation, one should supply initial data for each of its modes.
Since in spherical symmetry each mode evolves separately, it would
be sufficient to analyze the evolution of a single $l$-mode of the wave from
a given initial data.
We shall specify this initial data on two characteristic (null)
surfaces, as illustrated in figure \ref{fig1}\footnote{For a
characteristic initial value problem to be well posed, it suffices to
specify only the value of the field on the initial characteristic surfaces
(and not its derivatives).}.
Specifically, we shall consider initial data in the form of some compact
outgoing pulse, specified on the ingoing  null surface
$v=0$:\footnote{The choice
$v=0$ does not limit the generality of the initial setup, due to
the time translation invariance of the background geometry.}
\begin{equation} \label{eq6}
\left\{\begin{array}{l}
        \Psi^{l}(u=u_{0})=0 \\
        \Psi^{l}(v=0)=\Gamma^{l}(u)
        \end{array} ,
\right.
\end{equation}
where $\Gamma^{l}(u)$ is some function of a compact support
between retarded times $u=u_{0}$ and $u=u_{1}$ (for brevity we
henceforth usually suppress the $l$-dependence of the functions under
consideration).
We shall assume that $|u_{1}|>2R$, namely that
the support of the initial function $\Gamma(u)$ is completely outside the
shell.

The choice of compact ('localized') initial data proves to be
convenient for the purpose of calculation. It also becomes
useful when we later try to characterize the way in which the late
time behavior of the wave depends on the location of the initial
pulse. For that reason, it is instructive to consider the case
$|u_{0}|\gg u_{1}-u_{0}$, for which $u_{0}$ becomes a single parameter
describing the location of the initial pulse.

To complete the specification of the initial value problem, a boundary
condition should be set at the origin of coordinates.
From a physical point of view, a wave starting out of regular initial data
and traveling outside the shell,  should not develop any
singular behavior. Also, the field should remain regular (in fact,
continuous) at the shell itself, since the jump in the potential
function through this surface is finite in magnitude . Then, no singularity
is expected to occur during the free evolution of the scalar field in
the Minkowski interior of the shell.
Enforcing regularity on $\phi$ then automatically implies that the
revised wave function $\Psi=r\phi$ should vanish at the origin
$r_{*}=r=0$.
To rule out unphysical divergent solutions, we hence impose the
supplementary boundary condition
\begin{equation} \label{eq6a}
\Psi(r_{*}=0)=0.
\end{equation}

In appendix \ref{appA} we show that the evolution equation (\ref{eq4}),
together with the initial conditions
(\ref{eq6}) and the boundary condition
(\ref{eq6a}), establish a well defined characteristic initial value problem
for the scalar field, with a unique solution anywhere inside the light
cone of the initial data.

\section{The iterative expansion}\label{secIII}

Our goal is to calculate the late time behavior of $\Psi$ at null-infinity
(namely, for $v\rightarrow\infty$, $u\gg M$).
In what follows we introduce an analytic scheme to construct this solution.
This technique is based on the
{\em iterative expansion}, to be defined now.

First, define
\begin{equation} \label{eq14a}
V_{0}(r)\equiv \frac{l(l+1)}{4r_{*}^{2}}
\end{equation}
and
\begin{equation} \label{eq14b}
\delta V(r)\equiv V(r)-V_{0}(r),
\end{equation}
with the potential $V(r)$ given by Eq.\ (\ref{eq5}).

Then, consider a decomposition of the wave function,
\begin{equation} \label{eq14}
\Psi = \Psi_{0}+\Psi_{1}+\Psi_{2}+\cdots ,
\end{equation}
such that the components $\Psi_{N}$ obey the recursion formula
\begin{equation} \label{eq15}
\Psi_{N,uv}+V_{0}\Psi_{N}=\left\{\begin{array}{ll}
                       0                     & ,N=0 \\
                       -(\delta V)\Psi_{N-1} & ,N>0
                       \end{array}
             \right.
\end{equation}
and satisfy the initial conditions
\begin{equation} \label{eq16}
\begin {array}{llr}\Psi_{N}(u=u_{0}) & = & 0 \ \ \ \ (\forall N\geq 0)\ \\
                   \Psi_{N}(v=0)     & = &  \left\{\begin{array}{ll}
                                               \Gamma (u) & ,N=0 \\
                                               0          & ,N>0
                                                \end{array}
                                         \right.
\end{array}
\end{equation}
and the boundary conditions
\begin{equation} \label{eq16a}
\Psi_{N}(r_{*}=0)= 0 \ \ \ \ (\forall N\geq 0) .
\end{equation}

Formally summing Eqs.~(\ref{eq15},\ref{eq16},\ref{eq16a}) over $N$, we
recover Eqs.~(\ref{eq4},\ref{eq6},\ref{eq6a}) for the complete wave $\Psi$.
This suggests that if the sum (\ref{eq14}) converges, it should
yield the correct function $\Psi$.

Equations (\ref{eq15},\ref{eq16},\ref{eq16a}) constitute an infinite
hierarchy of
initial value problems for the functions $\Psi_{N}$. Each of the evolution
equations consists of a simple homogeneous part, which is the free evolution
equation in Minkowski's spacetime. In addition,
each of the functions $\Psi_{N}$ (excluding $\Psi_{0}$) has a source term
proportional to the previous function in the series.
Therefore, in principle, if the solution for $\Psi_{0}$ is found, and
the appropriate (time domain) Green's function is constructed, then we
should be able to solve for the functions $\Psi_{N}$ one by one, using the
standard Green's function method.

Analytic arguments, as well as numerical simulations (both to be presented
later)
strongly suggest that at
null-infinity the series (\ref{eq14}) {\em does} converge to the correct
function $\Psi$. Moreover, in the case $|u_{0}|\gg M$, which will concern us
here, we find (both numerically and analytically) that the late-time
behavior of $\Psi$ at null-infinity is well approximated by $\Psi_{1}$
(the corrections coming from $N\geq 2$ are smaller by a factor proportional
to $M/u_{0}$). Thus, in our scheme, understanding the behavior
$\Psi_{1}$ alone will suffice to determine the essential features of the
late-time dynamics at null-infinity.

In the sequel we derive an exact analytic expression for $\Psi_{0}$ and for
the Green's function , and use these results to calculate the late time
behavior of $\Psi_{1}$ at null-infinity. We then discuss the contributions
coming from higher orders of the iterative expansion.

\section{derivation of $\Psi_{0}$} \label{secIV}

By definition, $\Psi_{0}$ admits the homogeneous wave equation
\begin{equation} \label{eq17}
\Psi_{0,uv}+V_{0}(r_{*})\Psi_{0}=0,
\end{equation}
where $V_{0}$ is the purely centrifugal potential defined in
Eq.\ (\ref{eq14a}). This equation is supplemented by the initial conditions
\begin{equation} \label{eq18}
\Psi_{0}=\left\{\begin{array}{ll}
                      0 & ,u=u_{0} \\
                      \Gamma (u) & ,v=0
                      \end{array}
             \right. ,
\end{equation}
and by the boundary condition specified in Eq.\ (\ref{eq16a}).
(recall that $\Gamma$ is a function of compact support, representing the
initial pulse of scalar radiation.)

In fact, $\Psi_{0}$ is nothing but the solution of the
analogous wave evolution problem in {\em Minkowski} spacetime\footnote{
In this respect, the iterative decomposition is actually ``an expansion
of Schwarzschild about Minkowski''.}.
To see that, notice that inside the shell Eq.\ (\ref{eq17}) is invariant
under the transformation from the $r_{*},t_{*}$ coordinates to the usual
flat-space coordinates $r$ and $t$.
Outside the shell, however, only the {\em functional} form of
$\Psi_{0}(u,v)$ is the same as in Minkowski, while the dependence
of the characteristic coordinates $u,v$ on the flat space coordinates
$r,t$ differs.

In appendix \ref{appA} we show that a solution $\Psi_{0}$ to
Eq.\ (\ref{eq17}), subject to both the initial conditions (\ref{eq18}) and
the boundary condition at $r=0$, is {\em unique}.
This solution reads
\begin{mathletters} \label{eq20}
\begin{equation} \label{eq20a}
\Psi_{0}(u_{0}\leq u\leq 0)=\sum_{n=0}^{l} A_{n}^{l}
                          \frac{g_{0}^{(n)}(u)}{(v-u)^{l-n}}
\end{equation}
\begin{equation} \label{eq20b}
\Psi_{0}(u\geq 0)\equiv 0 ,
\end{equation}
\end{mathletters}
in which the coefficients $A_{n}^{l}$ are given by
\begin{equation} \label{eq21}
A_{n}^{l}=\frac{(2l-n)!}{n!(l-n)!}.
\end{equation}
The function $g_{0}(u)$ (with its parenthetical superindices indicating
the number of times this function is differentiated) is, by Eq.\
(\ref{eq18}), the solution to the inhomogeneous ordinary equation
of order $l$,
\begin{equation} \label{eq21b}
\sum_{n=0}^{l} A_{n}^{l}
\frac{g_{0}^{(n)}(u)}{(-u)^{l-n}}=\Gamma(u),
\end{equation}
subject to the initial conditions $g_{0}^{(n)}(u_{0})=0$ for all
$0\leq n\leq l-1$.
(Due to the compactness of $\Gamma (u)$, we then automatically have
also $g^{(l)}(u_{0})=0$. Thus $\Psi_{0}$ vanishes along the ray
$u=u_{0}$, as necessary.)

The solution for $g_{0}$ is given by
\begin{equation} \label{eq21a}
g_{0}(u\leq 0)=\frac{1}{(l-1)!}\int_{u_{0}}^{u}\!
\left(\frac{u}{u'}\right)^{l+1}(u-u')^{l-1}\Gamma(u')du'
\end{equation}
for $l\geq 1$, and simply $g_{0}(u)=\Gamma(u)$ for $l=0$\footnote{In the
scalar model, the monopole mode is radiative as any other mode.}.
It is easy to confirm this result by a direct substitution,
noticing that a
solution to the homogeneous equation corresponding to Eq.\
(\ref{eq21b}) has the general form
\begin{equation} \label{eq23}
g_{0}(u_{1}\leq u\leq 0)=\left\{\begin{array}{ll}
                        0                                & , l=0 \\
                        \sum_{k=1}^{l} \gamma_{k}u^{l+k} & , l>0
                      \end{array}
             \right. ,
\end{equation}
where $\gamma_{k}$ are constant coefficients.
Note that this is also the form of $g_{0}$ at $u_{1}\leq u\leq 0$
(for which retarded times Eq.\ (\ref{eq21b}) becomes
homogeneous), with the coefficients $\gamma_{k}$ being
certain functionals of the initial data function $\Gamma$,
that can be easily constructed by comparing Eqs.\ (\ref{eq21a}) and
(\ref{eq23}).

We find that during retarded times $u_{0}\leq u\leq u_{1}$, at
which the initial pulse is ``turn on'' on the initial ingoing null
surface, the behavior of $\Psi_{0}(u)$ depends on the detailed
structure of the pulse. At later (yet negative) retarded times,
$u_{1}\leq u\leq 0$, the function $g_{0}(u)$ takes the simple form
$u^{l+1}\times$ polynomial of order $l-1$ in $u$. Eq.\
(\ref{eq20a}) then implies that the wave $\Psi_{0}$ itself dies
off (as $\sim u$, generically) towards retarded time $u=0$.
Exceptional is the $l=0$ mode of the wave, which vanishes right
after the initial pulse `ceases' on the initial ray (that is, it
vanishes identically at $u\geq u_{1}$).

In the later zone of spacetime, $u>0$, the only solution for $\Psi_{0}$
satisfying both $\Psi_{0}(u=0)=0$\footnote{
Here we assert that the wave $\Psi_{0}$ is continuous through the ray
$u=0$. For, as can be easily verified, a discontinuity along this
ray must result is a violation of either the field equation or the
boundary condition. The continuity of the wave is also reasonable
from the physical point of view.}
and $\Psi_{0}(r_{*}=0)=0$, is the
trivial null solution (\ref{eq20b}) [see figure \ref{fig2}].
The uniqueness of this solution is guaranteed in virtue of the discussion
in Appendix \ref{appA}.

We can now unify both Eqs. (\ref{eq20}) by defining
\begin{equation} \label{eq26}
g_{0}(u>0)\equiv 0.
\end{equation}
It is also convenient to define $g_{0}(u<0)\equiv 0$, which makes $g_{0}(u)$
a function of compact support between $u=u_{0}$ and $u=0$.
Then, Eq.\ (\ref{eq20a}) suffices to describes the wave $\Psi_{0}$ anywhere.

\begin{figure}[htb]
\input{epsf}
\centerline{\epsfysize 6cm \epsfbox{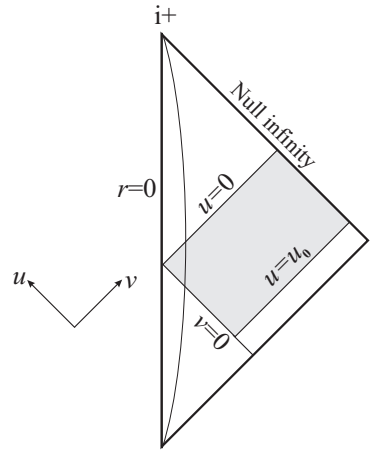}}
\caption{\protect\footnotesize The evolution of $\Psi_{0}$ (representing the
         scalar wave in Minkowski spacetime). For a compact initial pulse
         'turned on'
         during retarded times $u_{0}<u<u_{1}<0$, the support of
         $\Psi_{0}(u)$ is confined to $u_{0}<u<0$ (or $u_{0}<u<u_{1}$ for
         $l=0$). The region in which $\Psi_{0}$ has non-zero amplitude
         (for $l\geq 1)$ is represented by the shadowed area in this
         conformal diagram.}
\label{fig2}
\end{figure}

We conclude that for any mode of the scalar radiation, the evolution of
$\Psi_{0}$ is ``cut off'' not later then at $u=0$.
This somewhat surprising feature of the scattering
off the purely centrifugal potential originates, in the 1+1 representation,
from a destructive interference between ingoing
and outgoing wave fronts at the origin $r_{*}=0$
(see appendix \ref{appA} for details).
This may be more easily understood in the corresponding 3+1 picture, where
compactness of the initial pulse directly leads to compactness of the
wave fronts. The compact (in terms of retarded time) region in which
$\Psi_{0}$ survives is represented by the dark-colored area in the diagram of
figure \ref{fig2}.

Finally, we emphasize one further feature of $\Psi_{0}$.
By Eq.\ (\ref{eq20a}), we have {\em at null-infinity}
\begin{equation} \label{eq27}
\Psi_{0}^{\infty}(u)=g_{0}^{(l)}(u)
\end{equation}
in which $\Psi_{0}^{\infty}(u)$ stands for $\Psi_{0}(v\rightarrow \infty)$.
Now, the fact that $g_{0}(u)$ is compact with respect to retarded
time, means that any of its derivatives is compact as well.
We thus find that the $l$ integrals $(\Psi_{0}^{\infty})^{(-1)}$, \ldots ,
$(\Psi_{0}^{\infty})^{(-l)}$ , carried
out over all values of $u$ at null infinity, {\em all vanish}:
\begin{equation} \label{eq27a}
(\Psi_{0}^{\infty})^{(-n)}=\left. g_{0}^{(l-n)}
                           \right|_{-\infty}^{\infty}=0
\;\;\;\; \mbox{$(1\leq n\leq l)$}.
\end{equation}
This feature shall appear to have an important impact on the form of
the inverse power-law behavior of the scalar wave at late time.

Obviously, since $\Psi_{0}$ vanishes identically at $u>0$, it does not
contribute to the late time radiation. Rather, it serves as a source to
higher terms in our iterative expansion, as we show below.

\section{Construction of the Green's function} \label{secV}
In order to calculate the next terms in the iterative expansion (that is to
obtain solutions to the hierarchy of inhomogeneous equations (\ref{eq15}) for
$N\geq 1$), we shall use the standard Green's function approach.
To that end, we first need to obtain the Green's function corresponding to
the operator $\partial_{v} \partial_{u}+V_{0}^{l}$. This is the purpose of
this section.

The (retarded) Green's function $G(u,v;u',v')$ shall be defined as a solution
of the equation
\begin{equation} \label{eq28}
G_{,uv}+V_{0}^{l}G=\delta (u-u') \delta (v-v'),
\end{equation}
subject to the causality condition $G=0$ outside the future light cone
of the delta source point at $(u',v')$. In the sequel it will become
evident that this condition, together with an appropriate boundary condition,
specify a unique solution for $G$.

\begin{figure}[htb]
\input{epsf}
\centerline{\epsfysize 6cm \epsfbox{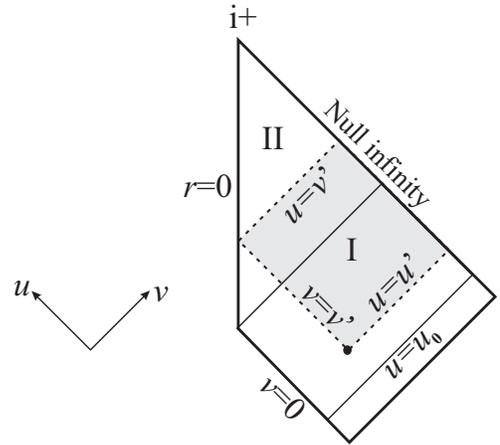}}
\caption{\protect\footnotesize The Green's function for a given source
          2-sphere
          at $(u',v')$. $G$ differs from zero only in region I
          (the shadowed area). It vanishes in region II.}
\label{fig3}
\end{figure}

To construct this solution we consider separately the two distinct regions of
spacetime indicated in figure \ref{fig3}, which are defined with respect to
given a source point (representing a source 2-sphere in 3+1 dimensions) at a
certain location $(u',v')$.
These are (I) the region inside the future light cone of $(u',v')$
not causally influenced by the origin (that is $v\geq v'$, $u'\leq
u\leq v'$);
and (II) the region inside the future light cone of $(u',v')$ which {\em is}
causally influenced by the origin (that is $u>v'$).

First consider the Green's function in region I. We impose causality by
writing
\begin{equation} \label{eq29}
G=\bar{G}\,\theta(u-u')\,\theta(v-v'),
\end{equation}
where $\bar{G}$ is a solution to the homogeneous equation (\ref{eq28}) at
$u>u'$ and $v>v'$, and $\theta$ is the standard step function.
By analogy to Eq.\ (\ref{eq20a}) we then have the solution
\begin{equation} \label{eq30}
\bar{G}(u,v;u',v')=\sum_{n=0}^{l} A_{n}^{l}\frac {g_{G}^{(n)}(u)}
{(v-u)^{l-n}},
\end{equation}
in which the coefficients $A_{n}^{l}$ are those given by Eq.\ (\ref{eq21}),
and where the function $g_{G}(u;u',v')$ is yet to be determined.
(Alternatively, we could have expressed this solution in terms of
a function $h_{G}(-v)$ instead of $g(u)$. Of course, the unique solution for
$G$, to be derived below, would have been the same).

Inserting Eq.\ (\ref{eq29}) into Eq.\ (\ref{eq28}), we find
that we must have
\begin{equation} \label{eq31}
\bar{G}(u=u')=\bar{G}(v=v')=1.
\end{equation}
With Eq.\ (\ref{eq30}), the equality $\bar{G}(v=v')=1$ takes the form of
Eq.\ (\ref{eq21b}), in which we first make the replacements
$g_{0}\rightarrow g_{G}$, $u\rightarrow (u-v')$ and $\Gamma(u)\rightarrow 1$.
Consequently, we find by Eq.\ (\ref{eq21a}) that (in region I)
\begin{equation} \label{eq32}
g_{G}(u;u',v')=\frac{1}{(l-1)!}\int_{u'}^{u}
               \left(\frac{u-v'}{u''-v'}\right)^{l+1}(u-u'')^{l-1}du''.
\end{equation}

To calculate the integral we make use of the formula
\begin{eqnarray} \label{eq32a}
\lefteqn{\int \frac{(x-x_{1})^{n_{1}}}{(x-x_{2})^{n_{2}}}dx=}  \nonumber\\
& &-\sum_{j=0}^{n_{1}}\frac {n_{1}!(n_{2}-j-2)!}{(n_{1}-j)!(n_{2}-1)!}
\frac {(x-x_{1})^{n_{1}-j}}{(x-x_{2})^{n_{2}-j-1}},
\end{eqnarray}
which is valid for any two natural numbers $n_{1}$ and $n_{2}$ satisfying
$n_{2}\geq n_{1}+2$.
(we shall need this integral several more times in the sequel).
With some additional algebraic manipulations we can then finally obtain
\begin{equation} \label{eq33}
g_{G}(u;u',v')=\frac{1}{l!}\left[\frac {(v'-u)(u-u')}{(v'-u')}\right]^{l}.
\end{equation}
It is easy to verify that with this result, we have also $G(u=u')=1$, as
necessary.

Next, consider region II, where
each of the functions $\Psi_{N}$ is subject to the boundary
condition $\Psi_{N}(r_{*}=0)=0$.
We impose this condition by requiring that the Green's function connecting
any source point to the origin should vanish:
\begin{equation} \label{eq34}
G(r_{*}=0;u',v')=0.
\end{equation}
Note that this boundary condition fails to be valid at the single
point $u=v=v'$, where its value, dictated by the evolution in region I, is
found by Eqs.\ (\ref{eq30}) and (\ref{eq32}) to be unity.
Nevertheless, the boundary condition for the various functions $\Psi_{N}$
is still satisfied at all values of $u$, as will become clear later.

In appendix \ref{appA} we show that, with the above boundary condition
imposed, the only solution to Eq.\ (\ref{eq28})
in region II is the null solution, namely $G(u>v')\equiv 0$.
We conclude that the Green's function is given by
\begin{eqnarray} \label{eq35}
G(u,v;u',v')& = & \sum_{n=0}^{l} A_{n}^{l}\frac {g_{G}^{(n)}(u;u',v')}
                  {(v-u)^{l-n}}                          \nonumber\\
            &\mbox{}\times & \theta(v-v')\theta(u-u')\theta(v'-u),
\end{eqnarray}
where the function $g_{G}$ is the one specified in
Eq.\ (\ref{eq33}).\footnote{
Note that the Green's function suffers a discontinuity along the
outgoing ray $u=v'$. This discontinuity, which is of order unity, originates
at the delta source point $(u',v')$ and travels along the ingoing ray $v=v'$.
When this discontinuity encounters the origin, it is reflected back along
$u=v'$. Some subtleties related with the appearance of this discontinuity
are discussed is appendix \ref{appA}.}

\begin{figure}[htb]
\input{epsf}
\centerline{\epsfysize 6cm \epsfbox{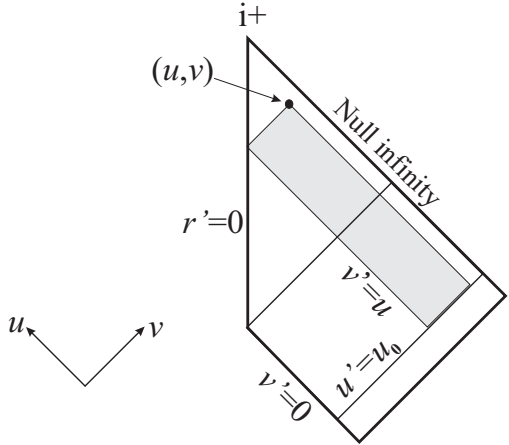}}
\caption{\protect\footnotesize The Green's function for a given evaluation
          point at $(u,v)$. The wave $\Psi_{N}$ at that point
          is influenced only by $\Psi_{N-1}$ sources inside the shadowed
          area.}
\label{fig4}
\end{figure}

In figure \ref{fig3} we have indicated the region of spacetime to which the
Green's function
$G(u,v;,u',v')$ carries the influence of a given source point located at
$(u',v')$.
In practice, however, we shall be interested in calculating the functions
$\Psi_{N}(u,v)$ (for
$N\geq 1$) at a certain evaluation point $(u,v)$. This will involve
integration over all point sources $(u',v')$.
The region in which sources influence the behavior of the wave at a given
evaluation point $(u,v)$ is indicated in figure \ref{fig4}.

\section{Calculation of $\Psi_{1}$ at null infinity} \label{secVI}

We have seen that the first element of our iterative series, namely
$\Psi_{0}$, do not contribute to the overall late time radiation.
Rather, it serves as a source to higher terms in the series.
The first contribution to the late time radiation comes from $\Psi_{1}$.
In what follows we shall give a detailed calculation of $\Psi_{1}$.
Specifically, we will obtain an analytic expression for $\Psi_{1}$ at
null-infinity, evaluated to the leading order in $M/u$. The
result will be highly significant, since this term will appear to be
the dominant constituent of the overall wave $\Psi$ at null infinity.
We shall discuss this issue later, when we analyze the higher ($N\geq 2$)
terms of the iterative expansion.

For $\Psi_{1}$ we have, by definition,
\begin{equation} \label{eq36}
\Psi_{1,uv}+V_{0}\Psi_{1}=-(\delta V)\Psi_{0},
\end{equation}
with the initial conditions $\Psi_{1}(u=u_{0})=\Psi_{1}(v=0)=0$,
and the boundary condition $\Psi_{1}(r_{*}=0)=0$.
The functions $V_{0}(r)$ and $\delta V(r)$ were given in
Eqs.~(\ref{eq14a}) and (\ref{eq14b}), respectively.

The solution can be formally written as
\begin{equation} \label{eq37}
\Psi_{1}(u,v)=
-\!\int_{u_{0}}^{u}\!\!\!\!du'\!\!\int_{u}^{v}\!\!\!\!dv'\,G(u,v;u',v')
 \delta V (u',v') \Psi_{0}(u',v')
\end{equation}
{[}with $\Psi_{0}(v'<0)\equiv 0$ understood],
where $G$ is the Green's function in Minkowski's spacetime, given in
Eq.\ (\ref{eq35}). This form manifestly admits the above
initial conditions. Also, the boundary condition is clearly
satisfied, as necessary. Notice that although
Eq.\ (\ref{eq34}) fails to hold at $u=v=v'$,
still the wave $\Psi_{1}$ vanishes at that point [because
the $v'$ integration in Eq.\ (\ref{eq37}) has no support in this
occasion.]

\begin{figure}[htb]
\input{epsf}
\centerline{\epsfysize 6cm \epsfbox{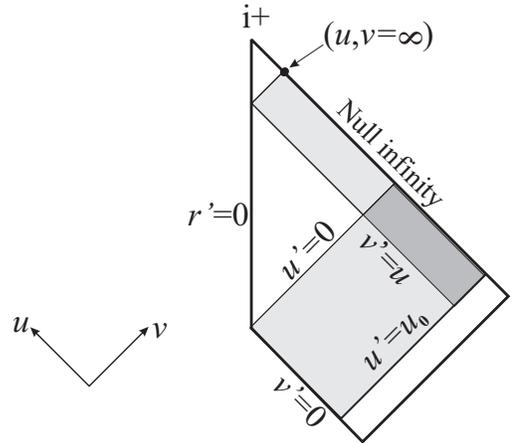}}
\caption{\protect\footnotesize Region where $\Psi_{0}$ sources affect
$\Psi_{1}$ at null infinity (indicated by the intersection of the two dark
colored areas). If $\Psi_{1}$ is evaluated at late retarded time ($u\gg M$)
then effective sources are confined to large distances ($r'\gg M$).}
\label{fig5}
\end{figure}

The region of integration in Eq.\ (\ref{eq37}), for a given evaluation
point $(u,v)$, is represented by the rectangle shown in figure \ref{fig4}.
Sources outside this rectangle do not influence the behavior of
$\Psi_{1}$ at $(u,v)$. In addition, we have found previously that the
support of $\Psi_{0}$
is confined to retarded times $u_{0}<u<0$. Hence, effective sources to
$\Psi_{1}$ are located only at the region indicated in
figure \ref{fig5} by the intersection of the two shadowed areas.
We thus observe that when evaluated at late retarded time $u$, $\Psi_{1}$
is influenced only by sources at large radii. As $u\rightarrow \infty$,
it is only the asymptotically-far region of spacetime that affects the
behavior of $\Psi_{1}$.

We shall now evaluate Eq.\ (\ref{eq37}), in order to yield an analytic
expression for $\Psi_{1}$ at null infinity, that is in the limit
$v\rightarrow \infty$. In this limit, the Green's function reduces to the
form
\begin{equation} \label{eq38}
G(u,v\rightarrow\infty;u',v')=\frac{\partial^{l}}{\partial u^{l}}
[g_{G}(u;u',v')].
\end{equation}
with only the $n=l$ term surviving in Eq.\ (\ref{eq35}), and where
$g_{G}$ is given by Eq.\ (\ref{eq33}).

We then notice that the $l$-derivative of $g_{G}$ with respect to $u$
can be taken out of the double integral in Eq.\ (\ref{eq37}).
The resulting surface terms all
vanish, due to the factors $(u-u')^{l}$ and $(v'-u)^{l}$ appearing in the
numerator of $g_{G}$ [see Eq.\ (\ref{eq33})].
Therefore, in analogy to Eq.\ (\ref{eq27}), we can now have
\begin{equation} \label{eq39}
\Psi_{1}^{\infty}(u)=\frac{\partial^{l}}{\partial u^{l}}[g_{1}^{\infty}(u)],
\end{equation}
where we define
\begin{equation} \label{eq40}
g_{1}(u,v)=-\!\int_{u_{0}}^{u}\!\!\!\!du'\!\!\int_{u}^{v}\!\!\!\!dv'\,
g_{G}(u;u',v') \delta V(u',v') \Psi_{0}(u',v'),
\end{equation}
and where $\Psi_{1}^{\infty}$ and $g_{1}^{\infty}$ stands for the value
of these function at null infinity ($v\rightarrow \infty$).
It would hence be sufficient to calculate $g_{1}^{\infty}$ in order to
immediately obtain $\Psi_{1}^{\infty}$.

To allow explicit integration in Eq.\ (\ref{eq40}), we must first express the
function $\delta V(r)$ in terms of the null coordinates.
This cannot be done explicitly outside the
shell, since the function $r(r_{*})$ is implicit at $r>R$. However, in terms
of the $M/r_{*}$ expansion we can write (Recalling that the parameter
$R$ is of order $M$)
\begin{equation} \label{eq41}
\delta V(r_{*}>r_{R})=\frac{a+b\ln 2\tilde{r}_{*}}{8r_{*}^{3}}
+ \O\left(\frac{M^{2}(\ln \tilde{r}_{*})^{2}}{r_{*}^{4}}\right),
\end{equation}
where $\tilde{r}_{*}\equiv r_{*}/(R-2M)$, $a$ is some constant
(depending on $R$ and $M$), and
\begin{equation} \label{eq42}
b=8Ml(l+1).
\end{equation}
Throughout the rest of this paper, the symbol ``$\sim$'', when appearing over
a character, shall always indicate the ratio of that quantity to $R-2M$.

Note that inside the shell we have $\delta V\equiv 0$  by definition,
and so the region $r_{*}'<r_{*}(R)=R f(R)^{-1/2}$ is automatically
excluded from the domain of integration in Eq.\ (\ref{eq40}).

The various terms in the $M/r_{*}$ expansion of $\delta V$ contribute to
$\Psi_{1}$ in an additive way [via Eq.\ (\ref{eq40})]. These contributions
may be, in principle, calculated for each of the terms in separate.
However, the calculation to follow suggests that to the leading order in
$M/u$, it is only the leading order of $\delta V$ in $M/r_{*}$ which
contributes to $\Psi_{1}$ at null infinity at late time . We shall therefore
focus now on the contribution coming from this leading order.

With the explicit form of $g_{G}$ and $\Psi_{0}$, and taking $\delta V$
to the leading order in $M/r_{*}$, Eq.\ (\ref{eq40}) reads at null
infinity
\begin{eqnarray} \label{eq43}
g_{1}^{\infty}(u) & = &
 -\frac{1}{l!}\sum_{n=0}^{l} A_{n}^{l}\!
\int_{u_{0}}^{u}\!\!\!\!du'g_{0}^{(n)}(u')\!\int_{u}^{\infty}\!\!\!\!\!\!dv'
\frac{(v'-u)^{l}(u-u')^{l}}{(v'-u')^{2l-n+3}}     \nonumber\\
 & & \times\ \left[a+b\ln (\tilde{v}'-\tilde{u}')\right]
 \theta [r_{*}'-r_{*}(R)],
\end{eqnarray}
where the fact that $\delta V=0$ inside the shell accounts for
the step function appearing in the integrand.

To proceed, we discuss separately the cases $l=0$ and $l\geq 1$.
First, consider {\bf the case $\bbox{l\geq 1}$}.
For these modes we show that the `$r_{*}^{-3}$ potential' source
(i.e.~the one proportional \mbox{to $a$}) contributes nothing to
$g_{1}^{\infty}$ at $u\geq 0$.
An analytic expression for the contribution coming from
the `logarithmic potential' source (the one proportional \mbox{to $b$})
shall then be derived.

To show that contribution from scattering off the $r_{*}^{-3}$
potential vanishes (in the case $l\geq 1)$, consider the term in
(\ref{eq43}) proportional to
$a$. Integrating over $v'$, with the help of Eq.\ (\ref{eq32a}),
this term yields
\begin{equation} \label{eq44}
a\sum_{n=0}^{l} B_{n}^{l} \int_{u_{0}}^{u}du'(u-u')^{n-2}
g_{0}^{(n)}(u'),
\end{equation}
with a vanishing contribution from the upper boundary of the
$v'$-integration, and where only the $j=l$ term has survived in
Eq.\ (\ref{eq32a}). The coefficients $B_{n}^{l}$ are given by
\begin{equation} \label{eq45}
B_{n}^{l}=\frac{l-n+1}{n!(2l-n+1)(2l-n+2)}.
\end{equation}

We now observe that in Eq.\ (\ref{eq44}) all terms corresponding to
$2\leq n\leq l$ vanish independently for $u\geq 0$, since $g_{0}(u)$ is
compact.
This can be verified by integrating each of the terms in the sum by
parts \mbox{$n-2$} successive times with respect to $u'$,
and then using Eq.\ (\ref{eq27a}) to find that all resulting surface terms
vanish.
Moreover, one also finds
that the two remaining terms of Eq.\ (\ref{eq44}), corresponding to $n=0$
and $n=1$, add to zero. This is easily shown by noticing that
$B_{0}^{l}=B_{1}^{l}$ for any $l\geq 1$.

We conclude that all terms proportional to $a$ in Eq.\ (\ref{eq43})
vanish (for $u\geq 0$), and thus that scattering off the $r_{*}^{-3}$
potential does not affect the late time behavior of the $l\geq 1$ modes
at null-infinity.
This seemingly odd feature becomes clear by the following simple argument.
Consider the solution $\Psi_{0}^{\epsilon}$ to the wave
equation (\ref{eq17}), in which we take the potential to be
$V_{0}^{\epsilon}=\frac{l(l+1)}{4(r_{*}+\epsilon)^{2}}$.
In addition, we take $\Psi_{0}^{\epsilon}$ to satisfy the same
initial conditions as $\Psi_{0}$, and the boundary condition
$\Psi_{0}^{\epsilon}(r_{*}=\epsilon)=0$.
Clearly, we have $\Psi_{0}^{\epsilon}(\epsilon\rightarrow 0)=\Psi_{0}$.
Differentiating the wave equation for $\Psi_{0}^{\epsilon}$ with
respect to $\epsilon$ and taking the limit $\epsilon\rightarrow 0$, we
obtain
\begin{equation} \label{eq45a}
\Psi'_{0,uv}+V_{0}\Psi_{0}'=\frac{l(l+1)}{2r_{*}^{3}}\Psi_{0},
\end{equation}
where $\Psi_{0}'\equiv \lim_{\epsilon\rightarrow 0}
\frac{\partial \Psi_{0}^{\epsilon}}{\partial \epsilon}$.
Comparison of Eq.\ (\ref{eq45a}) with the wave equation (\ref{eq36})
for $\Psi_{1}$, shows that $\Psi_{0}'$ should be proportional to
$\Psi_{1}$ calculated with $\delta V \sim r_{*}^{-3}$.
(Both functions, $\Psi_{1}$ and $\Psi_{0}'$, are subject to the
same initial conditions: on the initial surfaces we have, by definition,
$\Psi_{1}=0$ and also $\Psi_{0}'=0$, since the initial conditions
chosen for $\Psi_{0}^{\epsilon}$ are $\epsilon$-independent.)
Now, it is not difficult to verify that $\Psi_{0}'$ vanishes
identically at $u\geq 0$ (as does $\Psi_{0}$)\footnote{
                A straightforward calculation shows
                that $\Psi_{0}^{\epsilon}$ dies off at $u>2\epsilon$ as
                $\sim\exp[-u/(2\epsilon)]$. This implies the vanishing of
                $\partial \Psi_{0}^{\epsilon}/\partial \epsilon$
                at $u\geq 0$ in the limit $\epsilon\rightarrow 0$.}.
Therefore the contribution to $\Psi_{1}$ due to scattering off the
$r_{*}^{-3}$ potential must vanish identically at $u\geq 0$, as
demonstrated above for $\Psi_{1}^{\infty}$ by explicit
calculation.

Next, we have to consider contributions to $\Psi_{1}^{\infty}$
coming from scattering off the `logarithmic' potential
(i.e.\ contributions to the integral in Eq.\ (\ref{eq43}) due to terms
proportional to $b$). The details of this calculation are left to
appendix \ref{appB}, where we obtain the simple result
\begin{equation} \label{eq46}
g_{1}^{\infty}(u)=-2M\int_{u_{0}}^{0}\frac{g_{0}(u')}{(u-u')^{2}}du'.
\end{equation}

{\bf The case $\bbox{l=0}$ :}
The monopole case is especially simple to handle, as in this
case no logarithmic terms are involved in the calculation (the
coefficient $b$ in Eq.\ (\ref{eq41}) vanishes), and the Green's
function is simply unity.
For $l=0$ we have no summation over $n$ in Eq.\ (\ref{eq43}), and
have $a=4M$.
This equation then reads
\begin{equation} \label{eq46a}
g_{1}^{\infty}(u)=-4M\int_{u_{0}}^{0}du'\int_{u}^{\infty}dv'
\frac{g_{0}(u')}{(v'-u')^{3}} \mbox{\ \ (for $l=0$),}
\end{equation}
which directly leads to Eq.\ (\ref{eq46}).
Therefore Eq.\ (\ref{eq46}) is correct for any of the
modes. (Note the interesting result that $g_{1}^{\infty}$
depends on $l$ only through the explicit form of $g_{0}$).

Thus far we were approximating the potential $\delta V$ by its large $r_{*}$
form, indicated in Eq.\ (\ref{eq41}). In principle, to obtain the correct
expression for
$g_{1}^{\infty}$, the complete potential $\delta V$ should be considered.
Each of the terms in the $M/r_{*}$ expansion of the potential $\delta V$
contributes additively as a source to $g_{1}(u)$. These terms are all
proportional to $r_{*}^{-k_{2}}(\ln \tilde{r}_{*})^{k_{1}} $, where
$k_{1}$ and $k_{2}$ are natural numbers, satisfying $k_{2}\geq 3$ and
$0\leq k_{1}\leq k_{2}-2$.
Above we have given a complete analysis concerning the leading contribution,
namely the one corresponding to $k_{2}=3$. We can show, following
a completely analogous
analytic treatment, that the contribution from the above
general term of the $\delta V$ expansion (with $k_{2}>3$) is dominated at
late retarded time by
\begin{equation} \label{eq47}
g_{1}^{\infty}(u;k_{1},k_{2})\propto M^{k_{2}-2}
\int_{u_{0}}^{0} g_{0}(u')\frac
{\ln^{k_{1}}(\tilde{u}-\tilde{u}')}
{(u-u')^{k_{2}-1}}du'.
\end{equation}
(Note that the $k_{2}=3$ case is special, in that in this case there is no
logarithmic dependence involved in the resulting form of
$g_{1}^{\infty}$.)
We thus conclude that omitting the contributions from terms other than
the leading term in $\delta V$, has no effect on $g_{1}^{\infty}(u)$ to the
leading order in $M/u$ and in $u_{0}/u$.

Hence, in virtue of Eq.\ (\ref{eq46}) we
obtain, to the leading order in $M/u$ and in $u_{0}/u$,
\begin{equation} \label{eq48}
g_{1}^{\infty}(u\gg u_{0})  =  -2MI_{0}\;u^{-2},
\end{equation}
in which we have defined
\begin{equation} \label{eq49}
\int_{u_{0}}^{0}g_{0}(u')du'=\int_{-\infty}^{+\infty}\!\!\!g_{0}(u')du'
\equiv I_{0}.
\end{equation}
(The first equality is valid in view of the compactness of $g_{0}(u)$.)

By Eq.\ (\ref{eq39}) we can now finally obtain (to the leading order in
$M/u$ and in $u_{0}/u$)
\begin{equation} \label{eq50}
\Psi_{1}^{\infty}(u\gg u_{0})=2(-1)^{l+1}(l+1)!MI_{0}\;u^{-l-2}.
\end{equation}
Note that initial data information is manifested at null infinity at late
time only through the single integral $I_{0}$.
Also, note that in our approximation there is no reference whatsoever in the
form of $\Psi_{1}$ to the radius $R$ of the shell.

Both the power index of the radiation tail and the amplitude coefficient
deduced above, are the same as
those obtained by Gundlach, Price and Pullin (GPP)(Eqs.~(11) and (18) in
\cite {Gundlach94I}) for the late time behavior of the `complete' scalar
wave at null-infinity, $\Psi^{\infty}$, using a different approach.
Yet, our result regards only one element of the `complete' wave,
namely $\Psi_{1}^{\infty}$. To reveal the overall behavior of the wave we
must analyze each of the terms in the iterative scheme, then sum up
their contributions. Thus, in order for our result to coincide with that
of GPP, $\Psi_{2}^{\infty}$ and the higher terms in our expansion `must' be
negligible at null-infinity.
In the following chapter we argue that this, however, is not the case in
general. We
will show that when considering generic initial date, one finds
additional unnegligible contributions to $\Psi^{\infty}$ at late time.
These contributions, coming from $\Psi_{2}$, $\Psi_{3}, \ldots$, shall have
the form $M^{2}u_{0}^{-1}u^{-l-2}$, $M^{3}u_{0}^{-2}u^{-l-2}$, etc.
In this respect, the result of GPP is not strictly correct: the amplitude
calculated in \cite{Gundlach94I} should fail to represent the amplitude of
$\Psi^{\infty}$ for a generic initial pulse. However, as we indicate later,
the result of GPP, as well as the form of $\Psi_{1}^{\infty}$ calculated
above, approximates the `real' behavior of the scalar wave in the case
where
the initial compact pulse is confined to large radii, away from the highly
curved region of spacetime. Mathematically speaking, it will be argued that
the equality $\Psi_{1}=\Psi$ holds at null-infinity to only the leading order
in $M/u_{0}$.

It is interesting to mention that other previous attempts to
calculate the late--time behavior of scalar waves in Schwarzschild
\cite{Price72,Ching95,Leaver86,Andersson97}, all
yielded an expression proportional to the mass $M$, omitting possible
contributions from higher powers of $M$.
To our best knowledge, no analytic method has been proposed to enable
calculation of
the late time radiation resulting from initial date located at small radii,
where curvature is large.
In this respect, our iterative scheme is no exception. However, our
method {\it does} provide a formal mean to quantitatively explore
this aspect of the analysis,
which was somewhat overlooked in several previous works.
We shall further refer to this issue later in this paper.

The expression we obtained above for $\Psi_{1}^{\infty}$ [Eq.\ (\ref{eq50})]
corresponds to the case of an initial pulse which is compact with respect to
retarded time $u$. This
pulse represents a wide family of physically reasonable initial
data.
Yet another realistic initial setup involves the presence of a static field
outside some radius (say, the surface of a spherically collapsing object)
up to some moment of time (say, the onset of collapse).
The corresponding late time behavior of $\Psi_{1}$ at null infinity can be
easily inferred from our previous results, as follows.

The static flat-spacetime solution to the Klein--Gordon equation
(\ref{eq17}), regular at infinity, reads
$\Psi_{0}^{\rm stat}=\mu r^{-l}$, where $\mu$ is a constant representing
the strength of the initial static field. The corresponding $g_{0}$ function,
namely the solution to Eq.\ (\ref{eq21b}) with
$\Gamma(u)=\Psi_{0}^{\rm stat}(v=0)$,
is $g_{0}^{\rm stat}=2^{l}[l!/(2l)!]\mu$.
In order to calculate $\Psi_{1}$ in this case, we need only to substitute
for $g_{0}$ in Eq.\ (\ref{eq46}), while changing the range of integration
to $\int_{-\infty}^{u_{1}}du'$, $u_{1}$ being some retarded time after the
initial pulse ceases on the initial ingoing ray. Doing so, we
find, to the leading order in $M/u$ and in $u_{1}/u$,
\begin{equation} \label{eq51}
g_{1}^{\infty}(u)[{\rm stat]}= -2^{l+1}M[l!/(2l)!]\mu\, u^{-1}
\end{equation}
which leads to
\begin{equation} \label{eq52}
\Psi_{1}^{\infty}(u)[{\rm stat}]=(-2)^{l+1}\frac{(l!)^{2}}{(2l)!}M\mu\,
u^{-l-1}.
\end{equation}
This result is again identical with the one derived by GPP (Eqs.~(12) and
(18) in \cite{Gundlach94I}) for static initial data.

\section{Higher terms of the expansion} \label{secVII}

We now turn to evaluate the higher terms $(N\geq 2)$ of the iterative
expansion, each defined by Eqs.~(\ref{eq15}---\ref{eq16a}). These terms
may be formally constructed in an inductive way by means of the
`flat space' Green's function derived in sec.\ \ref{secV}:
\begin{eqnarray} \label{eq53}
\lefteqn{\Psi_{N}(u,v)=}              \nonumber\\
& & -\int_{u_{0}}^{u}du'\int_{u}^{v}\!dv'
G(u,v;u',v')\delta V(u',v')\Psi_{N-1}(u',v').
\end{eqnarray}
Manifestly, this solution satisfies the initial and boundary
conditions specified in Eqs.\ (\ref{eq16}) and (\ref{eq16a}),
respectively.

In analogy to the treatment of $\Psi_{1}$ [Eqs.~(\ref{eq39}),(\ref{eq40})],
we can write
\begin{equation} \label{eq54}
\Psi_{N}(u,v)=\frac{\partial^{l}}{\partial u^{l}}\left[g_{N}(u,v)\right],
\end{equation}
in which
\begin{equation} \label{eq55}
g_{N}(u,v)\!=\!-\!\int_{u_{0}}^{u}\!\!\!\!\!du'\!\!\int_{u}^{v}\!\!\!\!\!dv'
g_{G}(u;u',v')\delta V(u',v')\Psi_{N-1}(u',v')
\end{equation}
(with $\Psi_{N-1}(v'<0)\equiv 0$ for any $N\geq 1$).

With the explicit form of $g_{G}$ and $\delta V$, and
following the same integration-by-parts procedure as in deriving
Eq.\ (\ref{eq46}), the last equation can be put into the form
\begin{eqnarray} \label{eq56}
g_{N}(u,v)=\sum_{k=0}^{l} \int_{u_{0}}^{u}du'\int_{u}^{v}dv'
\frac{(v'-u)^{l} (u-u')^{l-k}}{(v'-u')^{2l-k+3}}        \nonumber\\
\times \left[a_{k}^{l}+b_{k}^{l}\ln (\tilde{v}'-\tilde{u}')\right]
g_{N-1}(u',v')\, \theta [r_{*}'-r_{*}(R)],
\end{eqnarray}
in which $a_{k}^{l}$ and $b_{k}^{l}$ are constant coefficients.

We observe that the analysis of $\Psi_{N}$ for $N\geq 2$ involves two
additional technical difficulties, which were not present in the calculation
of $\Psi_{1}$.
First, the source wave is no longer compact with respect to retarded time,
but rather it extends infinitely to the
future. This means that $\Psi_{N}^{\infty}$ will be influenced by
sources located at any value of $r$, not only at null infinity.
(However, as we discuss later, the small-$r$ sources seem not to affect the
asymptotic late time behavior at null infinity).
The second technical difficulty is the fact that for $N\geq 2$,
the source functions $g_{N-1}(u',v')$ are also $v'$-dependent. Thus, in
order to calculate $\Psi_{N}^{\infty}$ for $N\geq 2$, one must first be
provided with the form of $g_{N-1}$ at {\it any} value of $v$,
not merely at null infinity.

To proceed, we shall first refer specifically to the case $N=2$, in order to
demonstrate that {\em the dominant contribution to $\Psi_{N}^{\infty}$ at
late time comes from $\Psi_{N-1}$ sources at null infinity}\ ; Sources at
small
$r$-values do not affect the behavior of the wave at null infinity to the
leading order in $M/u$. Under this assumption, we shall than calculate this
dominant contribution, to obtain the late time decay pattern of each of the
functions $\Psi_{N}$ at null infinity.
Numerical support shall be presented.

\subsection*{calculation of $\Psi_{2}$}

\begin{figure}[htb]
\input{epsf}
\centerline{\epsfysize 6cm \epsfbox{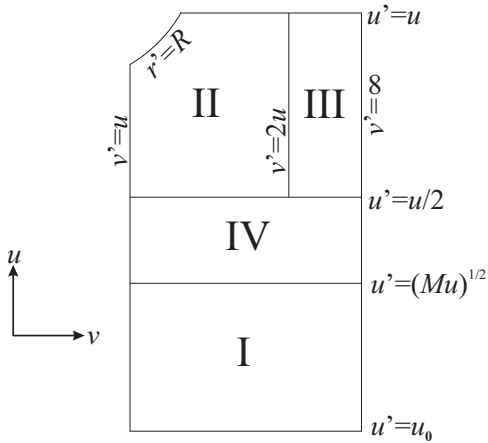}}
\caption{\protect\footnotesize Range of integration for
$\Psi_{2}^{\infty}(u)$.
The contribution from each of the regions I---IV, indicated in this sketch,
is evaluated in the text.}
\label{fig6}
\end{figure}

To analyze Eq.\ (\ref{eq56}) for $N=2$, we separate the domain
of integration into four regions, labeled I---IV, as indicated in figure
\ref{fig6}. In what follows we show that the dominant contribution to
$\Psi_{2}^{\infty}(u\gg M)$ comes only from sources in region I, which is
a distant
(large r value) region of spacetime. We shall thus call this zone `the main'
region of integration. An upper bound will be set on the contribution from
each of the other regions, II---IV, to verify their relative negligibility.

First, consider {\bf region II}, which is covered by $u\leq v'\leq 2u$ and
$u/2\leq u'\leq u$, with the sphere $r'<R$ excluded.
In this region of integration, the following upper bound is applicable to
the source function $g_{1}(u',v')$:
\begin{equation} \label{eq57}
|g_{1}(u',v')|<CM\frac{(v'-u')^{l+1}}{(u')^{l+3}}
\ln (\tilde{v}'-\tilde{u}_{0}),
\end{equation}
where $C$ is a positive constant.
To prove this, set $N=1$ in Eq.\ (\ref{eq56}), with the replacements
$(u,v)\rightarrow (u',v')$ and $(u',v')\rightarrow (u'',v'')$, and with the
$u''$ integration cut off at $u''=0$ [since $g_{0}(u''>0)=0$].
Then use the fact that for any $0\leq k\leq l$,
\begin{equation} \label{eq58}
\frac{(v''-u')^{l} (u'-u'')^{l-k}}{(v''-u'')^{2l-k+3}}\leq
\frac{(v''-u')^{l}}{(u'-u'')^{l+3}}\leq
\frac{(v''-u')^{l}}{(u')^{l+3}},
\end{equation}
where the first inequality arises since $v''\geq u'$,
and the second is due to the fact that $u''\leq 0$. The inequality
(\ref{eq57}) then follows from
$(\tilde{v}''-\tilde{u}'')\leq (\tilde{v}'-\tilde{u}_{0})$ and
after integrating over $v''$.

Now, to set an upper bound to the contribution from region II to
$\Psi_{2}^{\infty}$,
consider Eq.\ (\ref{eq56}) for $N=2$, with the double integral replaced
by $\int_{u/2}^{u}du' \int_{u}^{2u}dv'$
(see figure \ref{fig6}).
Using the inequality (\ref{eq57}), together with $(u-u')\leq (v'-u')$ and
$(v'-u)\leq (v'-u')$, one finds that
\begin{eqnarray} \label{eq59}
\lefteqn{\left|g_{2II}^{\infty}(u)\right| \leq}       \nonumber\\
& &{C_{1} M^{2}\ln^{2}\tilde{u} \int_{u/2}^{u}\!\!\!du'
\int_{{\rm max}\{u,u'+2R\}}^{2u}\!\!\!\!\!\!\!\!\!\!\!\!dv'
\frac{(v'-u')^{l-2}} {(u')^{l+3}}\leq}                \nonumber\\
& & C_{2}M^{2}u^{-3}\ln^{2}\tilde{u},
\end{eqnarray}
in which $g_{2II}^{\infty}(u)$ stands for the contribution from region II to
$g_{2}^{\infty}$, and where $C_{1}$ and $C_{2}$ are positive constants.

Next, consider {\bf region III}. The contribution $g_{2III}^{\infty}(u)$
arises from the integration \mbox{$\int_{u/2}^{u}du'\int_{2u}^{\infty}dv'$}.
The upper bound we have set above on $g_{1}(u',v')$ , Eq.\ (\ref{eq57}),
in no longer
efficient at large $r$ values. Here, we use a second upper bound, which may
be is easily derived from Eq.\ (\ref{eq56}):
\begin{equation} \label{eq60}
\left|g_{1}(u',v')\right|\leq C_{3}M(u')^{-2}
\ln \tilde{v}',
\end{equation}
where $C_{3}$ is yet another positive constant.
Provided with this upper bound, we observe that
\begin{eqnarray} \label{eq61}
\left|g_{2III}^{\infty}(u)\right|\leq C_{4} M^{2}
\int_{u/2}^{u}\!\!\!du' \int_{2u}^{\infty}\!\!dv' \frac{\ln^{2}\tilde{v}'}
{(u')^{2}(v'-u')^{3}}\leq           \nonumber\\
\leq C_{5}M^{2}u^{-3}\ln^{2}\tilde{u},
\end{eqnarray}
with $C_{4}$ and $C_{5}$ being positive constants.

We proceed by evaluating the contribution $g_{2IV}^{\infty}(u)$ from
{\bf region IV}. Using again the inequality (\ref{eq60}), we find that
\begin{eqnarray} \label{eq62}
\left|g_{2IV}^{\infty}(u)\right|\leq C_{6} M^{2}
\int_{\sqrt{uM}}^{u/2}\!\!\!\!\!du' \int_{u}^{\infty}\!\!dv'
\frac{\ln^{2}\tilde{v}'}{(u')^{2}(v'-u')^{3}}\leq           \nonumber\\
\leq C_{7}M^{2}u^{-5/2}\ln^{2}\tilde{u},
\end{eqnarray}
where $C_{6}$ and $C_{7}$ are still another positive constants.

Finally, we turn to calculate the contribution to $\Psi_{2}^{\infty}(u)$ due
to sources in {\bf region I}, that is at $u_{0}\leq u'\leq \sqrt{Mu}$.
To that end we write the source function $g_{1}$ in the form
\begin{equation} \label{eq63}
g_{1}(u',v')=g_{1}^{\infty}(u')+\Delta_{1}(u',v'),
\end{equation}
in which $\Delta_{1}(u',v')$ vanishes at null-infinity.
In appendix \ref{appC} we show that the contribution to $g_{2}^{\infty}$
due to this $v'$-dependent part of $g_{1}$ is bounded from above at large
retarded time by
\begin{equation} \label{eq64}
\left|g_{2\Delta}^{\infty}(u)\right|\leq
C_{8}M^{2}u^{-3}\ln^{2}\tilde{u}
\end{equation}
where $C_{8}$ is a positive constant.

It remains to calculate the contribution to $g_{2I}^{\infty}$ due to
$g_{1}^{\infty}(u)$.  With the source wave being a function of retarded
time only, the analysis is then exactly the same as the above analysis for
$\Psi_{1}^{\infty}$. In analogy to Eq.\ (\ref{eq48}), we thus immediately
obtain, to the leading order in $M/u$ and in $u_{0}/u$,
\begin{equation} \label{eq65}
g_{2I}^{\infty}=-2MI_{1}\;u^{-2},
\end{equation}
in which we have defined
\begin{equation} \label{eq66}
\int_{u_{0}}^{\sqrt{Mu}}g_{1}^{\infty}(u')du'\simeq
\int_{-\infty}^{\infty}g_{1}^{\infty}(u')du'\equiv I_{1}.
\end{equation}
The first equality, accurate to the leading order in $M/u$, stems from
the fact that the integrand is convergent at
$u\rightarrow \infty$ (since $g_{1}\sim u^{-2}$ at large $u$).

Collecting the above results [Eqs.~(\ref{eq59}), (\ref{eq61}), (\ref{eq62}),
(\ref{eq64}) and (\ref{eq65})] we conclude that
\begin{itemize}
\item The dominant contribution to $g_{2}^{\infty}(u\gg M)$ [hence also to
      $\Psi_{2}^{\infty}(u\gg M)$] is only due to sources located in the
      `main' region, namely at $r\gg M$; Sources at small $r$ values have a
      negligible effect on $g_{2}^{\infty}$ at late time.
\item To calculate the dominant contribution to $g_{2}^{\infty}(u\gg M)$,
      one is allowed to replace $g_{1}(u',v')$ in Eq.\ (\ref{eq56}) by
      $g_{1}^{\infty}(u)$. This does not affect $g_{2}^{\infty}$ to the
      leading order in $M/u$.
\end{itemize}
Therefore, Eq.\ (\ref{eq65}) approximates the `overall' late time behavior of
$g_{2}$ at null infinity. For the scalar wave itself we shall thus have,
by Eq.\ (\ref{eq54}),
\begin{equation} \label{eq67}
\Psi_{2}^{\infty}(u\gg m)=2(-1)^{l+1}(l+1)!MI_{1}\;u^{-l-2},
\end{equation}
which is analogous to Eq.\ (\ref{eq50}).

To summarize, we have found that $\Psi_{2}^{\infty}$ poses the same late time
behavior as $\Psi_{1}^{\infty}$, namely a $u^{-l-2}$ inverse power-law
decay (for the compact initial data setup). To find out the relative
amplitudes of these two terms at late time,
one must calculate the ratio of their coefficients, $I_{1}/I_{0}$.
This will be done in section \ref{secVIII}, after we first analyze the
behavior of the general term $\Psi_{N}$ of the iterative expansion.

\subsection*{The N'th term of the iterative expansion}
We would now like to understand the late time behavior of each of the higher
order terms ($N\geq 3$) of the iterative expansion. In principle, this may be
done in an inductive way, using Eq.\ (\ref{eq56}). We shall proceed by
{\em assuming} that, as was demonstrated above for $\Psi_{1}$ and $\Psi_{2}$,
the dominant contribution to $\Psi_{N}^{\infty}$ at late time
(for any $N\geq 1$) is only due to sources at null-infinity.
Phrased differently, we adopt the assumption that
{\it late time radiation at null-infinity originates predominantly from
scattering off spacetime curvature at null-infinity}. It means, in
particular, that it is only the asymptotically-far region of spacetime whose
structure affects the asymptotic late-time radiation at null infinity.

Accordingly, in order to obtain the late time form of $\Psi_{N}$ at
null-infinity, we will use Eq.\ (\ref{eq56}) with the source function
$g_{N-1}(u',v')$ replaced by $g_{N-1}^{\infty}(u')$, and with the upper limit
of the $u'$-integration set to $u'=\sqrt{uM}$. This will presumably provide
the correct form of $\Psi_{N}^{\infty}(u\gg M)$ to the leading order in
$M/u$, as was explicitly indicated in the case $N=2$.
The treatment of the $N\geq 3$ cases is then fully analogous to that
of $\Psi_{2}$, and the generalization of Eq.\ (\ref{eq65}) is straightforward:
\begin{equation} \label{eq71}
g_{N}^{\infty}(u\gg M)=-2MI_{N-1}\;u^{-2},
\end{equation}
in which we define
\begin{equation} \label{eq72}
I_{N-1}\equiv \int_{-\infty}^{\infty} g_{N-1}^{\infty}(u')du',
\end{equation}
and which is accurate to the leading order in $M/u$ and in $u_{0}/u$.
By Eq.\ (\ref{eq54}) this finally leads to
\begin{equation} \label{eq72a}
\Psi_{N}^{\infty}(u\gg m)=2(-1)^{l+1}(l+1)!MI_{N-1}\;u^{-l-2},
\end{equation}
which generalizes Eqs.~(\ref{eq50}) and (\ref{eq67}).

\begin{figure}[htb]
\input{epsf}
\centerline{\epsfysize 4.5cm \epsfbox{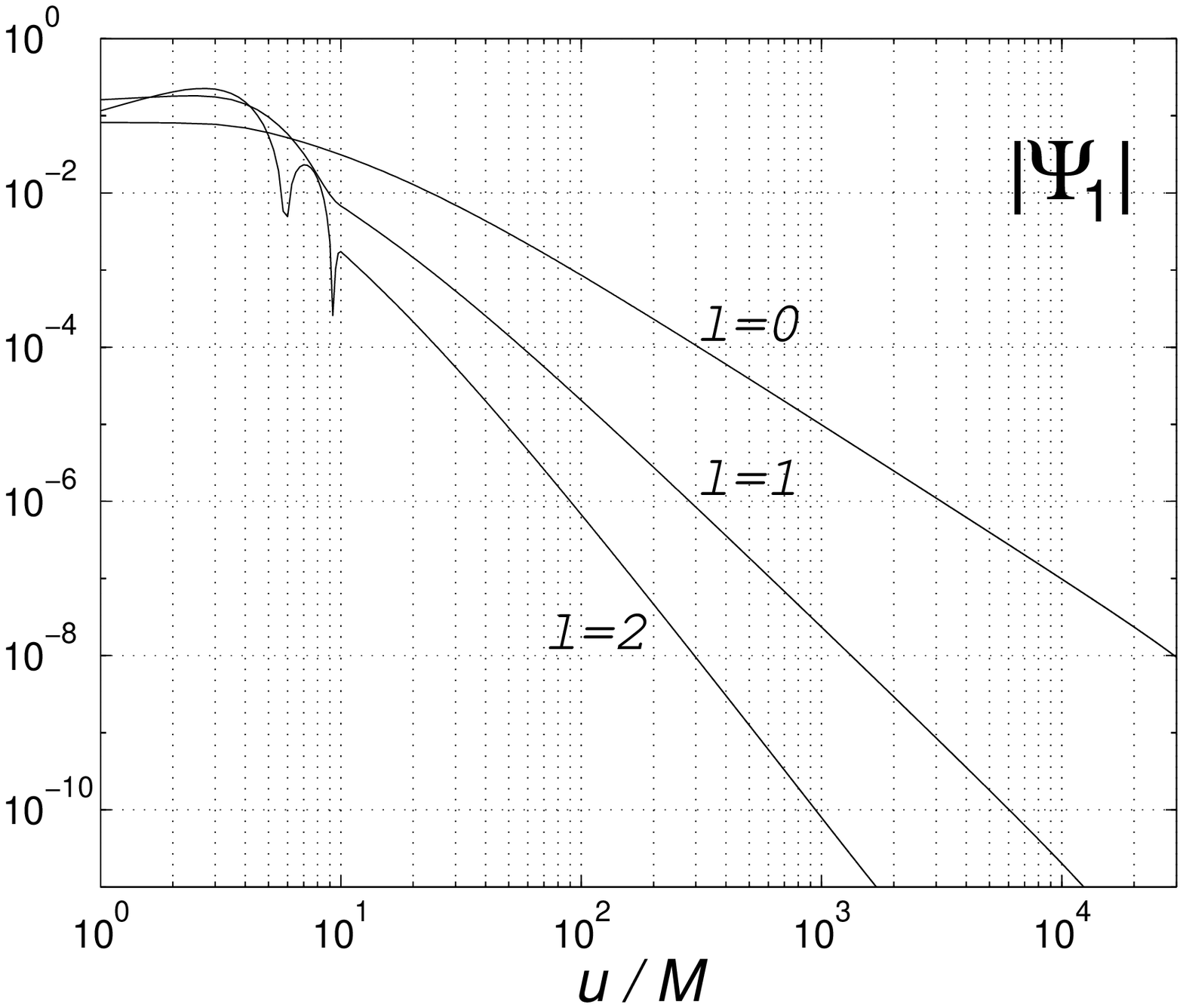}}
\centerline{\epsfysize 4.5cm \epsfbox{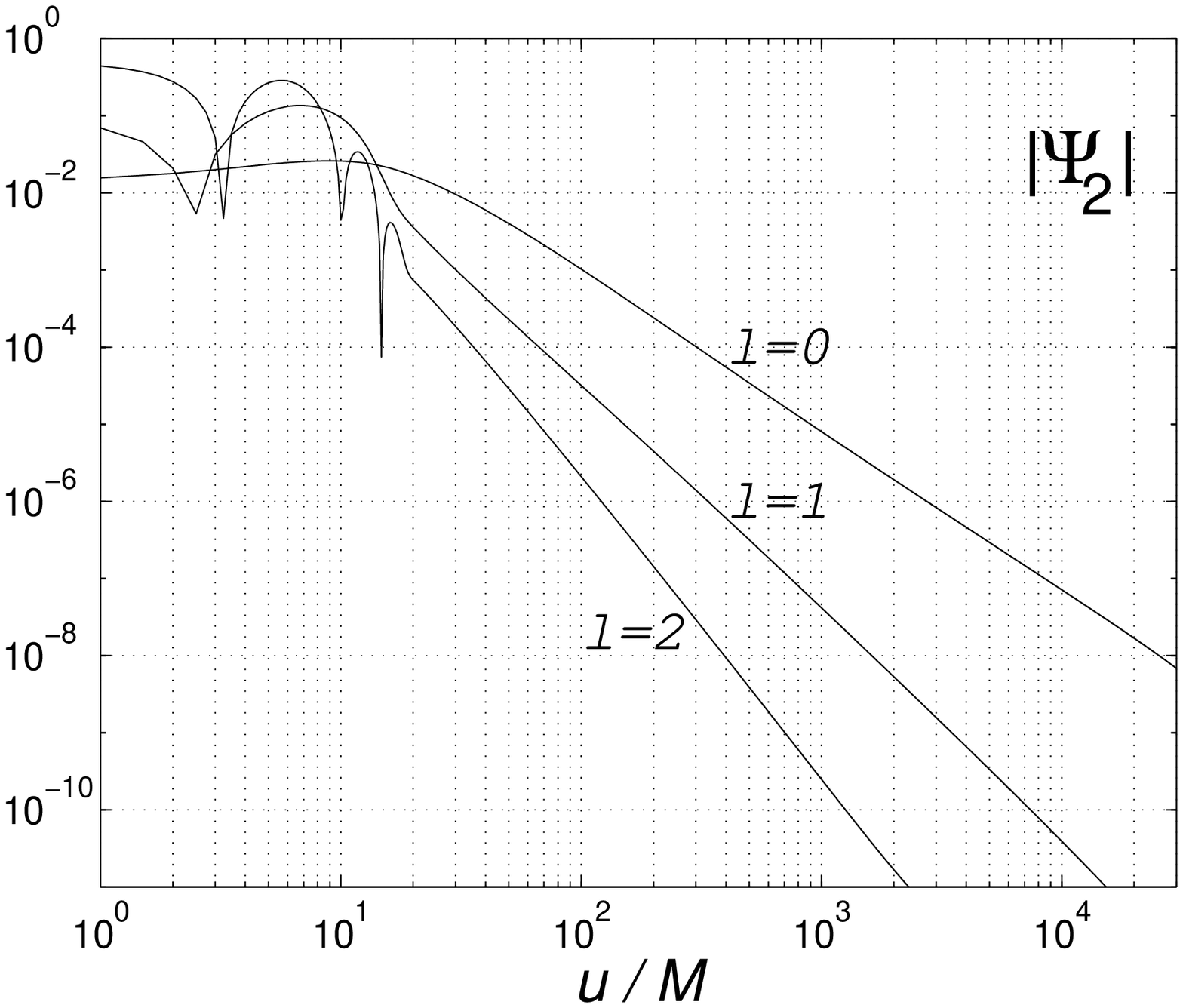}}
\centerline{\epsfysize 4.5cm \epsfbox{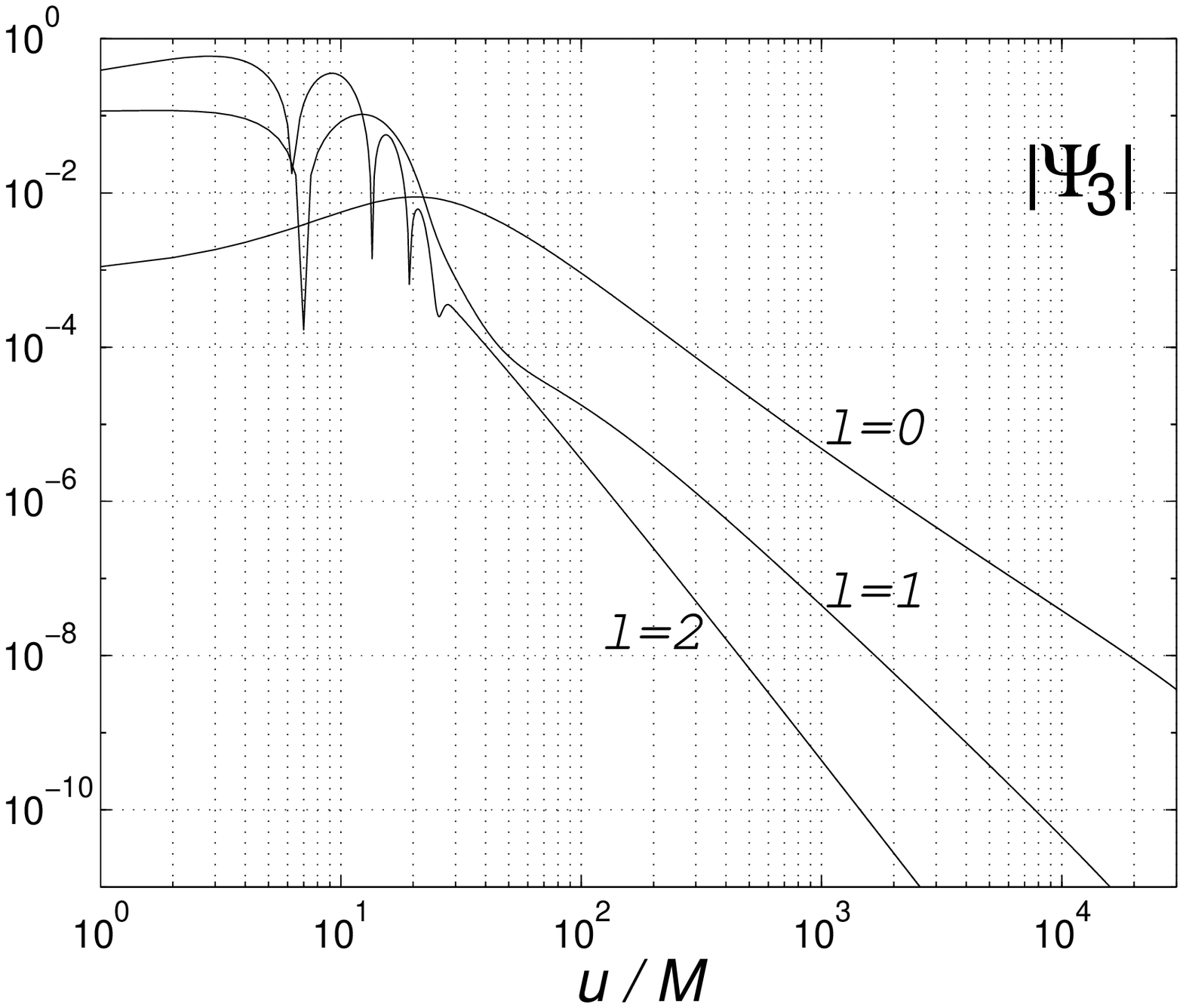}}
\caption{\protect\footnotesize Late time tails of the iterative
expansion terms. Presented on a log-log scale are numerical results obtained
for $\Psi_{1}$, $\Psi_{2}$ and $\Psi_{3}$  at $v=80000M$ (approximating
null infinity), for the $l=0,1,2$ modes.
Compact initial data for the propagation has been specified
between $u=-5M$ and $u=-10M$, and the radius of the shell has been set to
$R=3M$. The results demonstrate the $u^{-l-2}$ late time decay rate
predicted by the analytic calculation.}
\label{fig7}
\end{figure}

We conclude that each of the terms $\Psi_{N\geq 1}$ in the iterative
series has a similar late time behavior at null infinity, that is a
$u^{-l-2}$ inverse-power law decay. Numerical experiments firmly
support this observation, as demonstrated in figure \ref{fig7}.
Consequently, one finds that, in general, to obtain the correct overall
amplitude of the scalar wave at null infinity (even merely to the leading
order in $M/u$), one must sum all
contributions from the various terms $\Psi_{N\geq 1}$.
If this sum turns out to converge---then the overall power-law
should be $u^{-l-2}$.

A considerable simplification is achieved when the case $|u_{0}|\gg M$ is
considered. This corresponds to initial setup in which the pulse is
specified where spacetime is approximately flat (far outside the highly
curved region). In this case, analytic arguments as well as numerical
experiments, both to be presented below, suggest that (i) the iterative
series converges at null infinity at late time, and that (ii) the
dominating term of the
expansion there is $\Psi_{1}$, which well approximates the ``complete''
wave $\Psi$.  (The contribution from the rest of the terms is smaller by
order $M/u_{0}$).

\section{Convergence of the iterative expansion} \label{secVIII}

To discuss the convergence properties of the iterative expansion, we
shall concentrate on the case $|u_{0}|\gg M$. In what follows we argue that
the
ratio of any two successive terms of the iterative expansion, when evaluated
at null infinity at $u\gg M$, is given to the leading order in $M/u_{0}$ by
\begin{equation} \label{eq73}
\frac{\Psi_{N+1}^{\infty}}{\Psi_{N}^{\infty}}
\propto M|u_{0}|^{-1}\ln |\tilde{u}_{0}|
\end{equation}
(for $N\geq 1$) with a proportion factor depending on $N$.
Below we present numerical results in a firm
support of this relation. But first, we show that the validity of the ratio
(\ref{eq73}) is also suggested using analytic considerations.

Throughout the following discussion we shall refer to the limit
$R/u_{0}\rightarrow 0$ for simplicity. To that end we need to
assume that the limit $\Psi_{N}^{\infty}(R/u_{0}\rightarrow 0)$ exist, and
that the functions $\Psi_{N}^{\infty}$ are each independent of $R$ to
the leading order in $M/u$ and in $M/u_{0}$ (this is supported by
numerical examination).

To evaluate the ratio $\Psi_{N+1}^{\infty}/\Psi_{N}^{\infty}$ we shall have
to figure out how the
various functions $\Psi_{N}$ scale with respect to the parameter $|u_{0}|$.
To that end we first show that each of the functions
$g_{N}(u,v)$ can be written in the form
\begin{equation} \label{eq73a}
g_{N}(u,v)=u_{0}^{l-1-N}(\ln |\tilde{u}_{0}|)^{N}
f_{N}(\bar{u},\bar{v}),
\end{equation}
where $f_{N}$ are certain functions of only the dimensionless
variables $\bar{u}\equiv u/u_{0}$ and $\bar{v}\equiv v/u_{0}$.
This form, which is valid globally (for any values of $u$ and $v$) to the
leading order in $M/u_{0}$, may
be verified using mathematical induction: By Eq.~(\ref{eq21a})
we learn that Eq.\ (\ref{eq73a}) holds for
$g_{0}$.\footnote{To explore the scaling of $g_{0}$ with respect to $u_{0}$,
                  it is convenient to eliminate any scale parameters which
                  may be characteristic of the initial data function
                  $\Gamma(u)$, by referring specifically to the case
                  $\Gamma(u)=\delta(u-u_{0})$. Provided that any possible
                  scale characteristic of the initial date is much smaller
                  then $|u_{0}|$, this cannot limit the generality of our
                  discussion.}
Following the inductive procedure, we now assume that (\ref{eq73a}) is
valid for a certain value of $N$. We
then obtain for $g_{N+1}$ [using Eq.\ (\ref{eq56}) with $x\equiv u'/u_{0}$
and $y\equiv v'/u_{0}$],
\begin{eqnarray} \label{eq73b}
\lefteqn{g_{N+1}(u,v)=}                                         \nonumber\\
& & u_{0}^{l-2-N}(\ln |\tilde{u}_{0}|)^{N}
  \sum_{k=0}^{l}
  \int_{1}^{\bar{u}}\!\!dx \!\int_{{\rm max}(0,\bar{u})}^{\bar{v}}
  \!\!\!\!\!\!\!\!\!dy
\frac{(y-\bar{u})^{l}(\bar{u}-x)^{l-k}}{(y-x)^{2l-k+3}}     \nonumber\\
& &\times \left\{a_{k}+b_{k}\ln [\tilde{u}_{0}(y-x)]
\right\}f_{N}(x,y).
\end{eqnarray}
For $|u_{0}|\gg M$ this may be written (neglecting terms which are
smaller by $\sim\ln |\tilde{u}_{0}|$) as
\begin{eqnarray} \label{eq73c}
g_{N+1}(u,v)=u_{0}^{l-2-N}(\ln |\tilde{u}_{0}|)^{N+1}
f_{N+1}(\bar{u},\bar{v}),
\end{eqnarray}
where we have defined
\begin{equation} \label{eq73d}
f_{N+1}(\bar{u},\bar{v})\equiv \sum_{k=0}^{l} b_{k}
\int_{1}^{\bar{u}}dx \int_{{\rm max}(0,\bar{u})}^{\bar{v}}\!\!\!\!\!dy
\frac{(\bar{u}-y)^{l}(\bar{u}-x)^{l-k}}{(y-x)^{2l-k+3}}.
\end{equation}
Thus, by induction, Eq.\ (\ref{eq73a}) holds.

In virtue of Eqs.\ (\ref{eq73a}) and (\ref{eq71}) we then find
that (for $N\geq 1$)
\begin{equation} \label{eq73e}
I_{N-1}\propto u_{0}^{l+1-N}(\ln|\tilde{u_{0}}|)^{N},
\end{equation}
which by Eq.\ (\ref{eq72a}) describes the scaling of the functions
$\Psi_{N}^{\infty}(u\gg M)$
with respect to $u_{0}$. The ratio (\ref{eq73}) then follows
immediately.
[We comment that for the $l=0$ mode, no logarithmic factor is
expected to occur neither in Eq.\ (\ref{eq73}) nor in Eq.\
(\ref{eq73e})].

Eq.\ (\ref{eq73}) suggests that the
{\em late time behavior of the scalar wave at null infinity is dominated by
$\Psi_{1}^{\infty}$}, provided that $|u_{0}|\gg M$.
More accurately, it indicates that to the leading order in $M/u$, in
$u_{0}/u$ and in $M/u_{0}$ we have
\begin{equation} \label{eq74}
\Psi^{\infty}=\Psi_{1}^{\infty}.
\end{equation}
The significance of this result stems from the fact
that we have a simple analytic expression for $\Psi_{1}^{\infty}$
at late time.

\begin{figure}[htb]
\input{epsf}
\centerline{\epsfysize 5cm \epsfbox{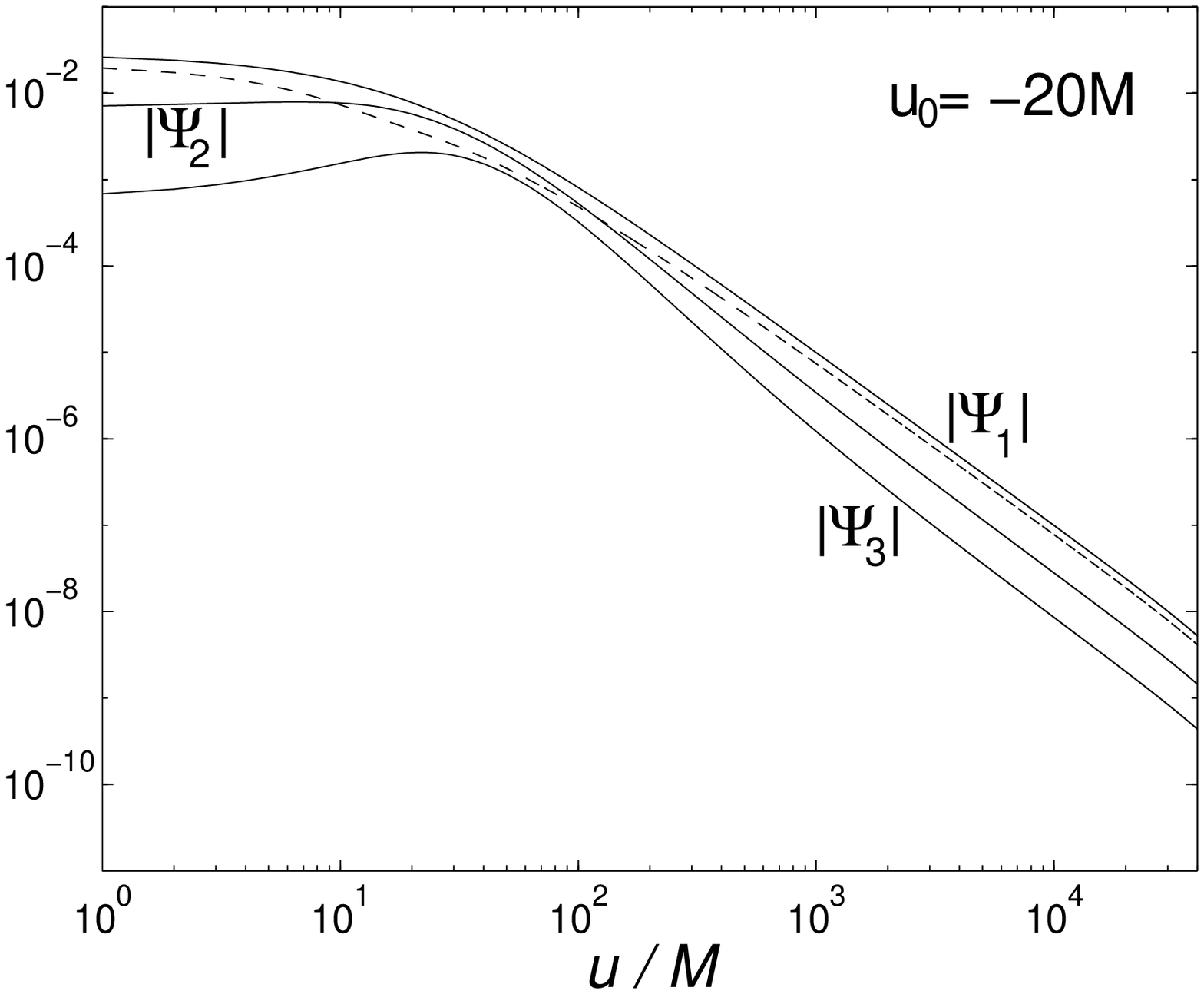}}
\centerline{\epsfysize 5cm \epsfbox{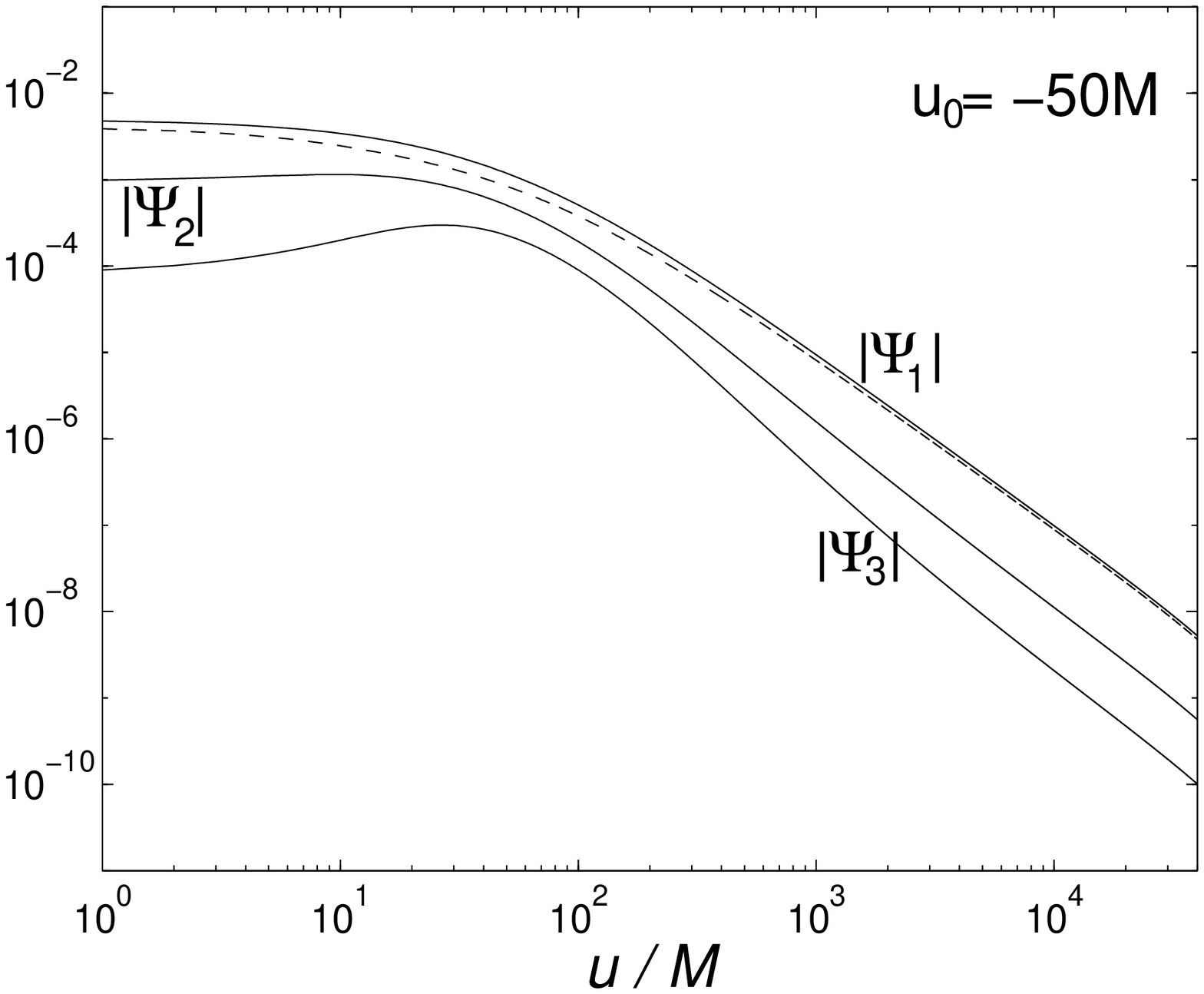}}
\centerline{\epsfysize 5cm \epsfbox{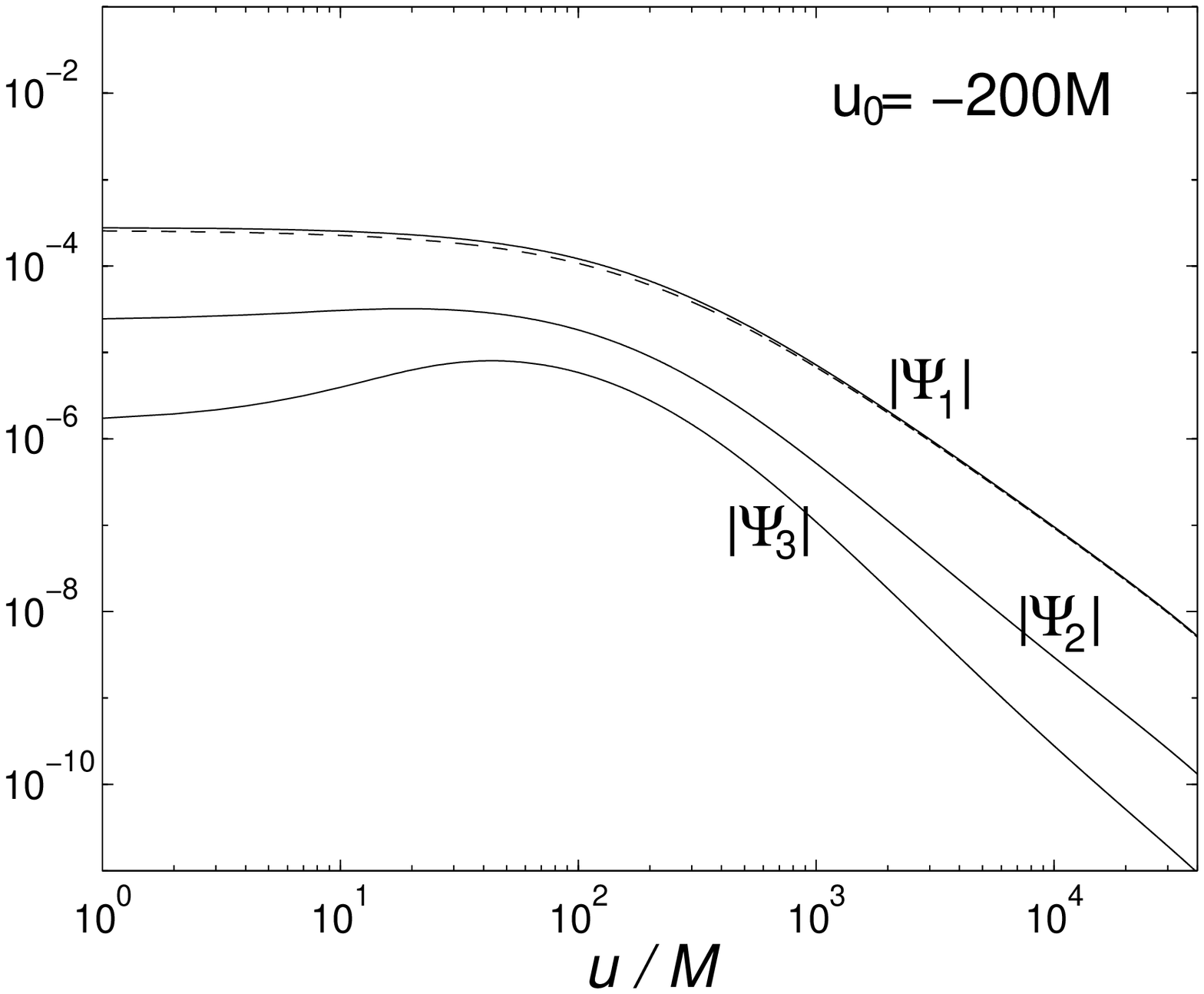}}
\caption{\protect\footnotesize Relative amplitudes of the
functions $\Psi_{N}$ at null infinity, for various values of the
parameter $u_{0}$. Shown are numerical results describing the $l=0$
mode of $\Psi_{1}$, $\Psi_{2}$ and $\Psi_{3}$ at $v=10^{5}M$
(approximating null infinity), for the cases
$u_{0}=-20M,-50M,-200M$. Also shown for reference (in dashed line) is the
 `complete'
wave $\Psi$, obtained by a direct numerical solution of the wave
equation (\ref{eq4}). The results support the observation that the
ratio $|\Psi_{N}/\Psi_{N+1}|$ decreases as $|u_{0}|/M$ is chosen
larger, and that for $|u_{0}|/M$ large enough, $\Psi_{1}^{\infty}$
becomes the dominant component of the `complete' scalar wave
$\Psi$, as suggested by analytic considerations.
Note the independence of the amplitude of $\Psi_{1}$ in the value
of $u_{0}$ at late time, which agrees with the scaling rule
(\ref{eq73e}) for the monopole mode (where no logarithmic factor is
involved).}
\label{fig8}
\end{figure}

\begin{figure}[htb]
\input{epsf}
\centerline{\epsfysize 5cm \epsfbox{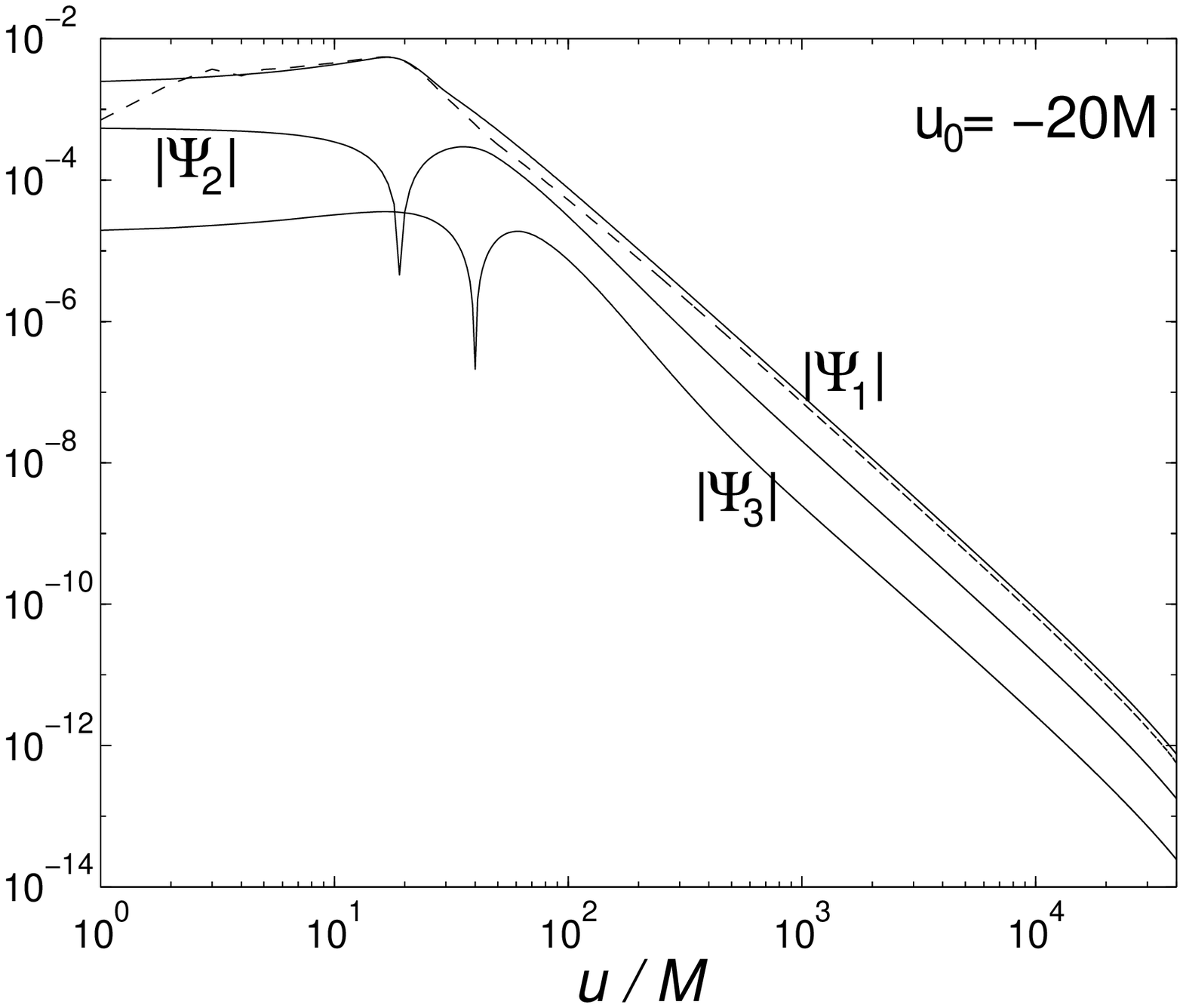}}
\centerline{\epsfysize 5cm \epsfbox{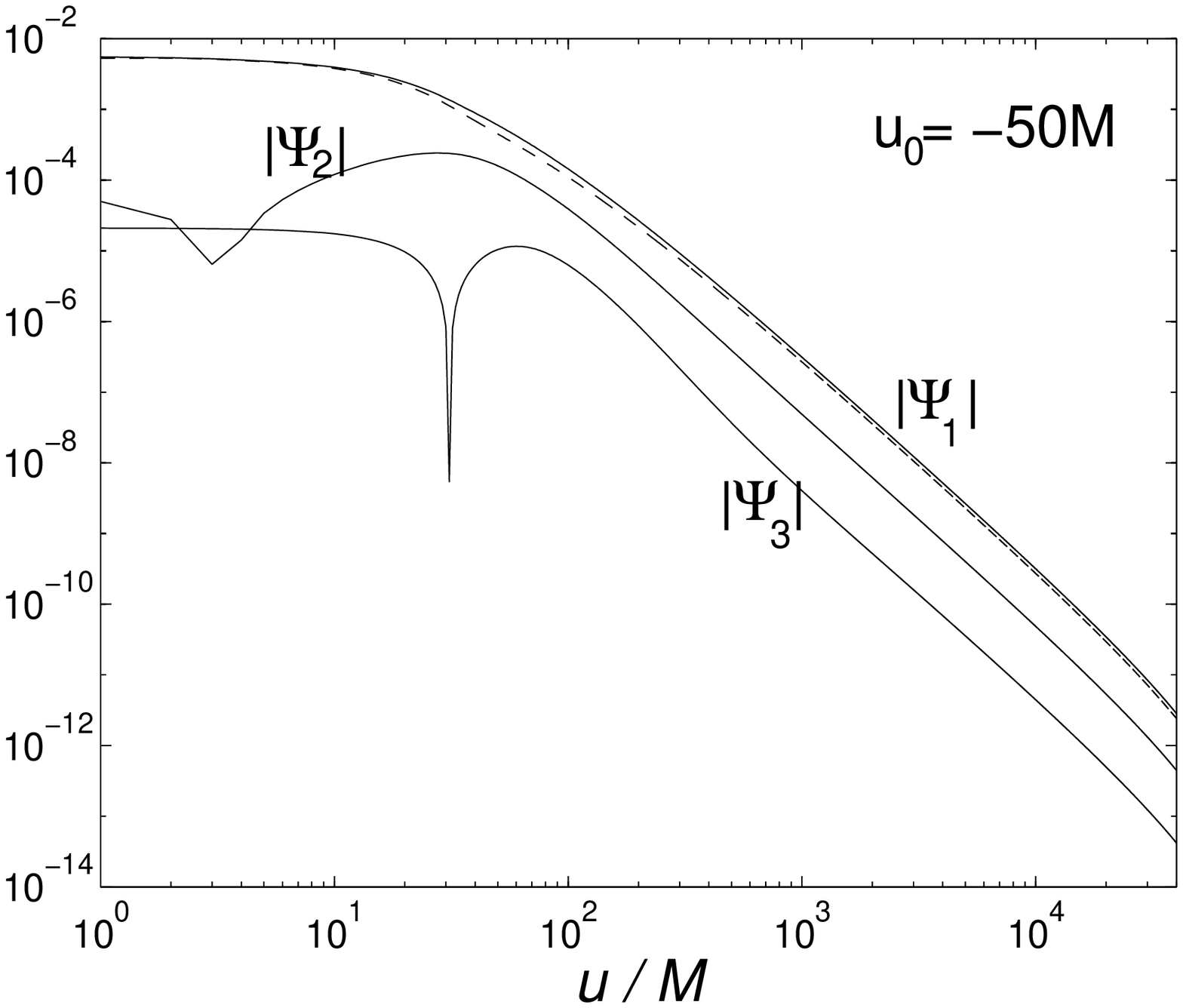}}
\centerline{\epsfysize 5cm \epsfbox{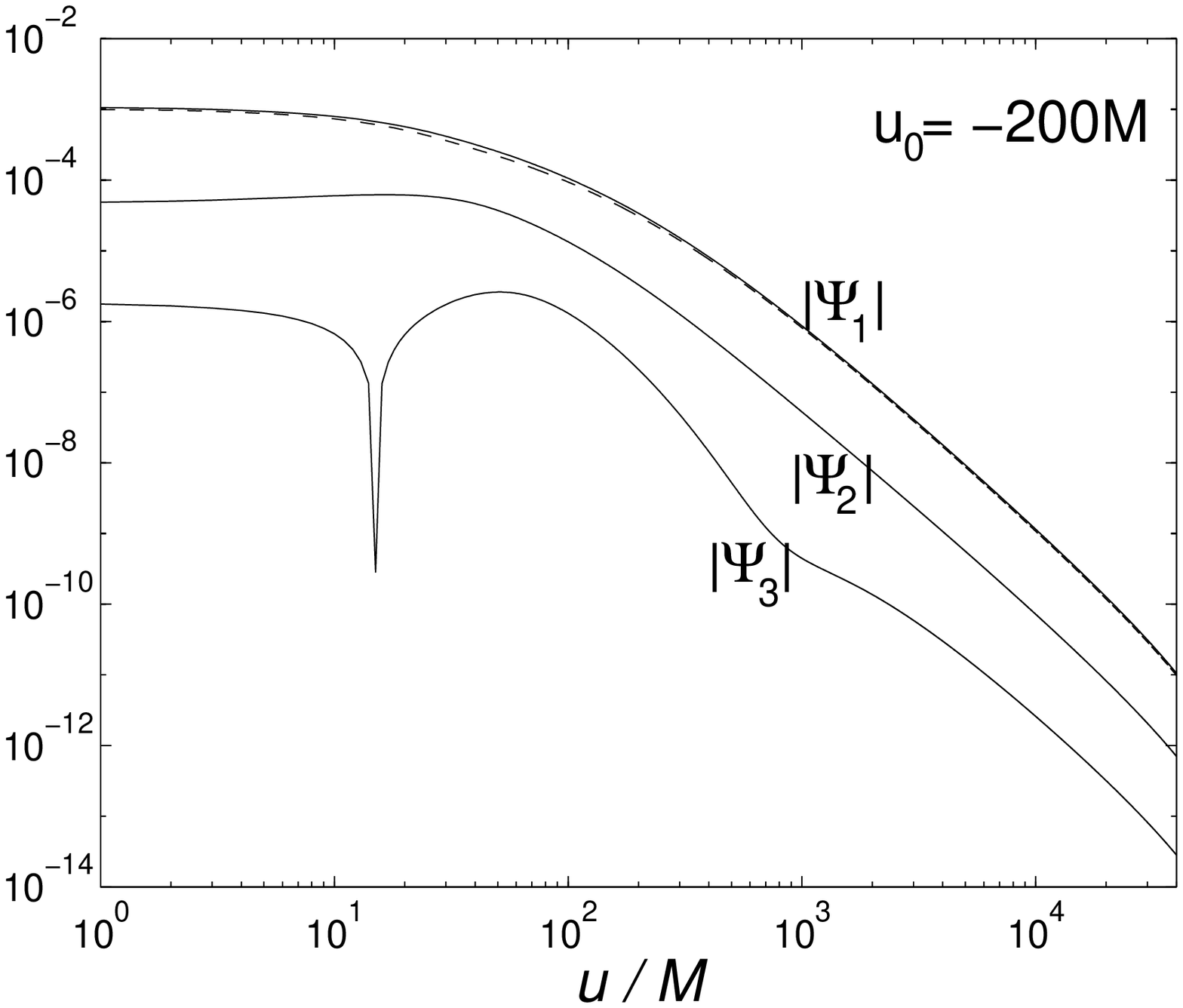}}
\caption{\protect\footnotesize The same as in figure \ref{fig8},
this time for the $l=1$ mode of the scalar field.
The steep decreases in the amplitude of $\Psi_{3}$ occur where
this function changes its sign (see the comment in the text).
In the case $u_{0}=-200M$, a ``remnant'' of such a sign-changing
at $u\simeq 1000M$ ``postpones''
the development of the power-law tail to later times.
}
\label{fig9}
\end{figure}

\begin{figure}[htb]
\input{epsf}
\centerline{\epsfysize 6cm \epsfbox{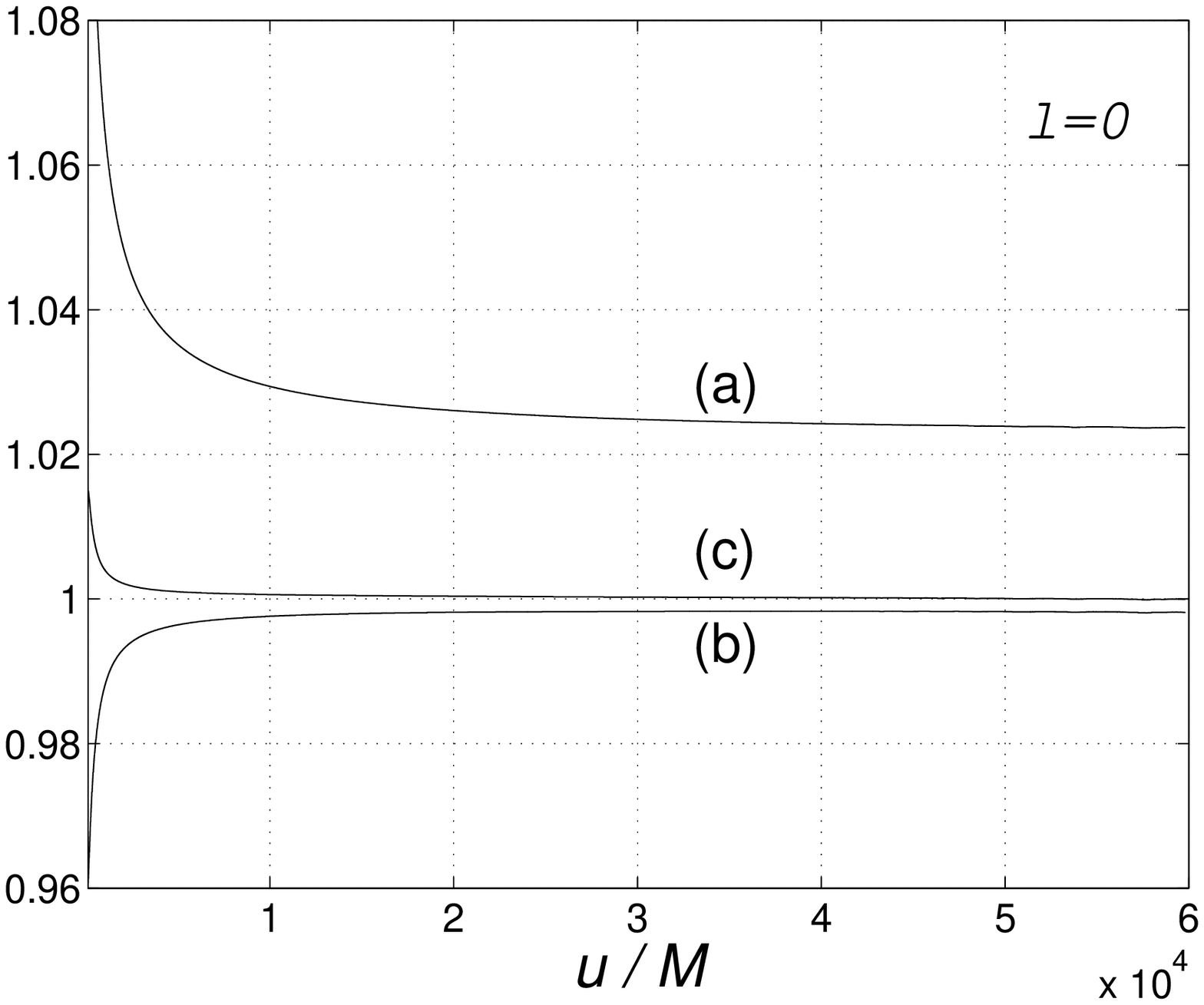}}
\centerline{\epsfysize 6cm \epsfbox{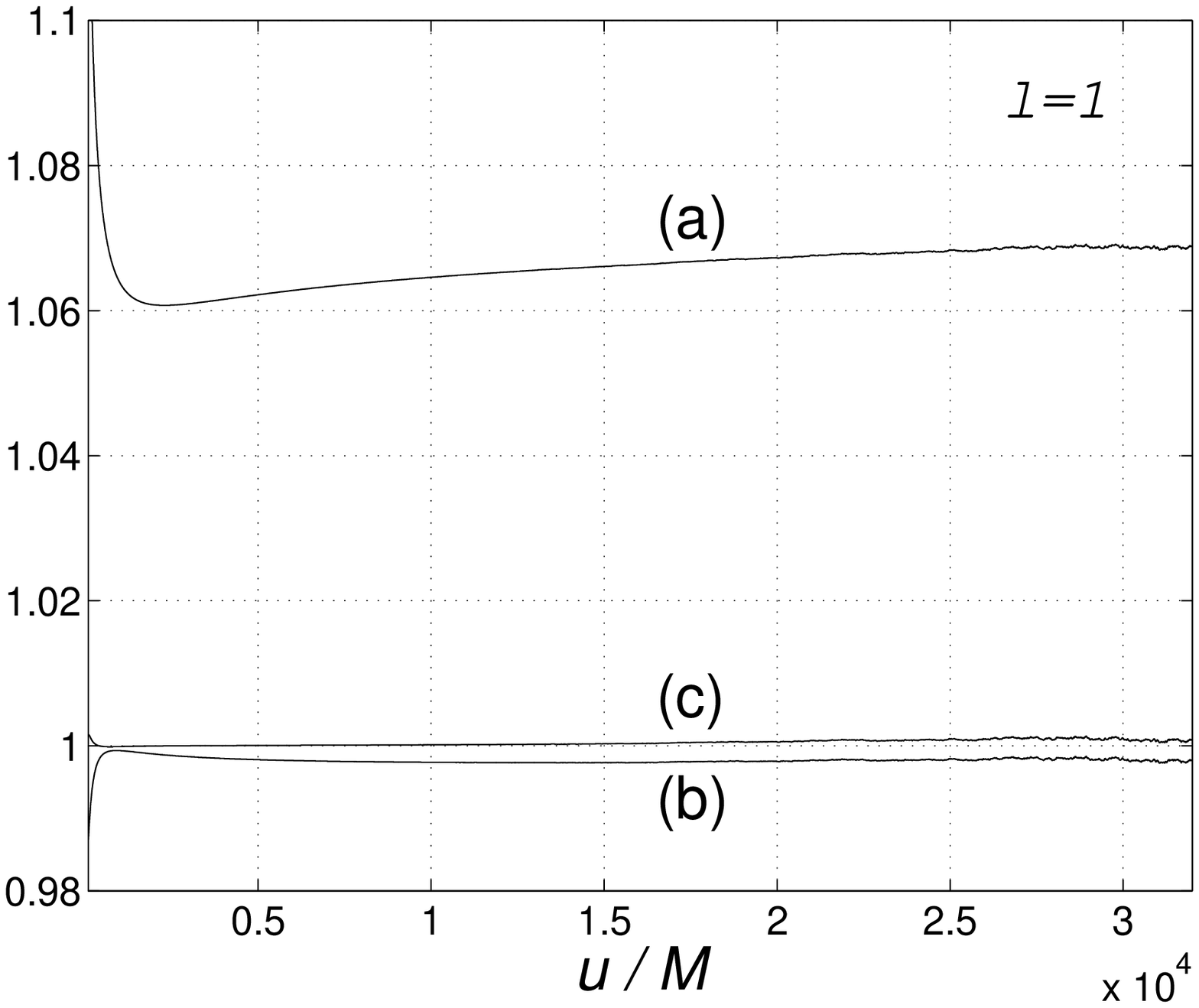}}
\caption{\protect\footnotesize Indications for convergence of the
iterative scheme. Presented (on a linear scale) are the ratios (a)
$\Psi_{1}/\Psi$, (b) $\frac{\Psi_{1}+\Psi_{2}}{\Psi}$, and (c)
$\frac{\Psi_{1}+\Psi_{2}+\Psi_{3}}{\Psi}$ for $l=0$ and for $l=1$.
Other parameters are $R=5M$,
$u_{0}=-200M$ and $v=10^{5}M$ (approximating null infinity).
The results demonstrate the convergence of the iterative expansion
for large $|u_{0}|/M$ values at large retarded time $u$.
}
\label{fig10}
\end{figure}

Numerical investigation, regarding the behavior of the functions
$\Psi_{N}^{\infty}$
as related to the value of $u_{0}$, provides a firm (though
qualitative) support to both Eqs.\ (\ref{eq73}) and (\ref{eq74}).
Examples of these numerical results are presented in figures
\ref{fig8}, \ref{fig9}, and \ref{fig10}.

Note that Eq.\ (\ref{eq73}) by itself does not tell us whether the sum
of the iterative expansion is convergent, as we do not know the form of
the ($N$-dependent) proportion factor appearing in this expression.
However, the numerical experiments strongly suggest that the expansion
{\em does} indeed converge, provided that $|u_{0}|/M$ is large enough.
Also, the rate of convergence seems to increase as we take $|u_{0}|/M$ to
be larger.
These features are also apparent in figures \ref{fig8} through \ref{fig10}.

\newpage

\newpage

The amplitude of the various functions $\Psi_{N}$ may change its sign
during the early stage of evolution (as apparent in figures
\ref{fig7} and \ref{fig9}). Our numerical experiments suggest that
this kind of sign-changing occur at larger and larger values of $u$
as $N$ and $u_{0}$ increase, ``delaying'' the formation of power
law tails to later and later times. We could not rule out the
(somewhat bothering)
possibility that for any value of $u$ and $u_{0}$, there will
exist some $N_{0}$ such that for any $N\geq N_{0}$ sign-changing would
occur at retarded time greater than $u$. In that case, our analytic
considerations [especially the ones leading to Eq.\ (\ref{eq73})]
will not apply for all $N$. Nevertheless, numerical examination
shows that the convergence itself is not disturbed by such sign-changing.

\begin{figure}[htb]
\input{epsf}
\centerline{\epsfysize 6cm \epsfbox{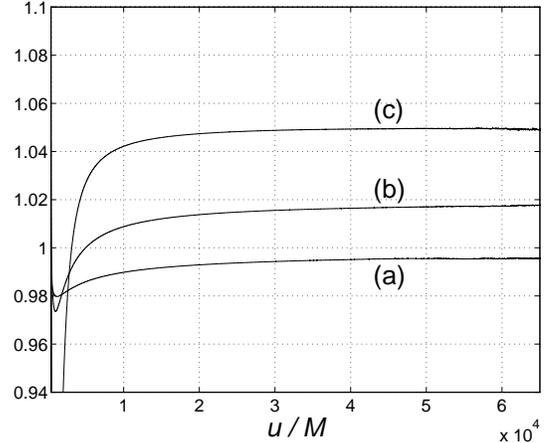}}
\caption{\protect\footnotesize Contribution from higher order
terms in $\delta V$ to $\Psi_{N}^{\infty}$ is negligible at $u\gg M$.
This is demonstrated here for the $l=1$ mode in the case $R=3M$ and
$u_{0}=-200M$. Plotted are the ratios (a) $\Psi_{1}/\Psi_{1}^{*}$,
(b) $\Psi_{2}/\Psi_{2}^{*}$, and (c) $\Psi_{3}/\Psi_{3}^{*}$, where the
functions $\Psi_{N}$ are those defined with the `complete'
potential $\delta V$, and the functions $\Psi_{N}^{*}$ are defined
with $\delta V$ taken to the leading order in $M/r_{*}$.
}
\label{fig11}
\end{figure}

Finally, a word of caution is in place:
In our analytic calculations we were taking into account only the
contributions to the functions $\Psi_{N}$ due to scattering off
the '$r_{*}^{-3}\ln r_{*}$'
potential, which is the asymptotic form of $\delta V$ at large distance.
We have argued previously that contributions due to the other terms in the
$M/r_{*}$ expansion of $\delta V$ do not affect the behavior at null infinity
at large $u$.
This, indeed, is supported numerically, as demonstrated in figure
\ref{fig11}.
However, such terms (each giving an additive contribution to
each of the functions $\Psi_{N}$) {\em do} affect the exact amplitude
of the leading-order tail: For example, these terms give an additional
contribution to the amplitude of $\Psi_{2}^{\infty}$, which (in terms of the
$M/u_{0}$ expansion) is of the same order as the amplitude of
$\Psi_{3}^{\infty}$. Therefore,
to examine whether the sum of the iterative series converges to the actual
'complete' wave $\Psi^{\infty}$ (at $u\gg M$), one must take into account
the 'complete' potential $\delta V$. (This, indeed, was done in obtaining the
results presented in figures \ref{fig8} through \ref{fig10}).

\section{summary and discussion} \label{secIX}

The problem of scalar wave evolution in a curved spherically-symmetric
spacetime can be formalized mathematically as a two-dimensional
initial-value problem, with a non-trivial effective potential. In this
paper we incorporated the simple shell model in order to demonstrate
the applicability, and explore the features, of a new calculation scheme
to handle this problem. This new scheme is based on a special perturbative
decomposition of the
wave (``{\em the iterative expansion}'') which, if fact, converts the
original homogeneous non-trivial initial-value problem into an infinite
hierarchy of inhomogeneous initial-value problems, each having a simple
Minkowski-like effective potential. The resulting initial-value problems
can then be treated analytically in an inductive manner, using the
Green's function in Minkowski spacetime.

Unfortunately,  we could not give a full analytic treatment concerning
the convergence properties of our scheme. However, some analytic
considerations, firmly supported by numerical analysis, indicated
that the role of the ``small parameter'' of the iterative
expansion is played by $M/u_{0}$ (although this parameter is not
self-manifested in the wave equations).
It was strongly suggested that the
iterative expansion {\em does} converge at null infinity, provided that
the support of the initial data
for the wave evolution is confined to large distances (small $M/|u_{0}|$).
Moreover, our analysis indicated that under the above condition, the late
time behavior of the `complete' wave is well approximated at null infinity
by $\Psi_{1}$, namely the first element of the expansion manifested at late
time. The last result is of special significance, since the late time form
of $\Psi_{1}$ at null infinity can be derived easily, as shown in section
\ref{secVI}.
Hence, for $|u_{0}|\gg m$ our iterative expansion proves to be a simple
effective tool for analyzing the late time behavior of the scalar field
at null infinity.

Here, it is important to distinguish between the issue of {\em
efficiency} of the scheme to that of its {\em usefulness}.
The iterative scheme
can still prove useful even when $M/|u_{0}|$ is too large for the
expansion to converge efficiently (that is, for $\Psi^{\infty}$ to
be well approximated by merely $\Psi_{1}$): In all cases when the scheme
converges, it always predicts (by Eq.\ (\ref{eq72a})) that the
overall late time tail of $\Psi^{\infty}$ has the form $u^{-l-2}$
(for the compact initial set-up). The convergence may be
`efficient' (when $|u_{0}|/M$ is large enough), allowing one to
obtain a simple analytic approximation for $\Psi^{\infty}$ by
calculating $\Psi_{1}$; or else it may be `inefficient', in which
case our scheme still provides a simple formal way to calculate
$\Psi^{\infty}$, by successive applications of the formula
(\ref{eq53}) (e.\ g.\ using numerical methods).

We think the following is a reasonable physical explanation
to the formal relation discovered between the smallness of the
parameter $M/|u_{0}|$ (describing the distance at which the initial
perturbation originates), and the effectiveness of the iterative expansion.
As already argued by Price \cite{Price72},
spacetime curvature at small distances acts as an
(almost) impenetrable potential barrier for the long waves
of the scalar field.
These long waves are the ones dominating the late time behavior of
the field (the effective frequency goes to zero as $t$ goes to
infinity).
Since long waves coming from large distance cannot reach the
highly-curved region (these waves are effectively scattered off the
potential tail at large distances), their evolution takes place
in the large-$r$ region of spacetime.
Thus, for a perturbation originating at large distance ($|u_{0}|\gg M$),
late time behavior of the field should depend mainly on the structure of
spacetime at large distance; details of spacetime curvature at
the highly-curved (small $r$) region should be less relevant for
this behavior.
This may no longer be true if the initial perturbation is
specified at small $r$, thus allowed to evolve through the
highly-curved region.
Now, the iterative scheme has the mathematical structure of a perturbative
expansion, with convergence properties that depend on how much
can $\delta V$ be considered a ``small'' amount.
This function gets arbitrarily small as $r$ gets arbitrarily
large, but at small distances it is finite, and not small.
However, the above reasoning implies that the form of $\delta V$
at small distance seems to be irrelevant to the wave
behavior at late time, provided that the initial perturbation
originates at large distance ($|u_{0}|\gg M$).
The above argument supplies a physical explanation to our formal
result that the effectiveness of the iterative scheme increases
as $|u_{0}|/M$ is taken larger.

In this paper we have focused on analyzing the wave at {\em null
infinity}.
The results we obtained have a clear physical significance with respect
to a distant static observer:
Along world-lines of constant $r$ we have
$\Delta u=\Delta t$, where $\Delta u$ and $\Delta t$ are, respectively,
the retarded time and the (flat-space) static observe's time elapsed
since the 'main pulse' of radiation had reached the observer.
Now, for $r$ very large, `null-infinity' is in fact the region
$\Delta t\ll r$ \footnote{The spacetime region $\Delta t<<r$,
for $r$ very large, may be properly named (after Leaver \cite{Leaver86})
'the {\em astrophysical zone}' of the waves.}.
Thus, in accordance with our results at null infinity, at times
$M\ll \Delta t\ll r$
such an observer observes a $\Psi\propto(\Delta t)^{-l-2}$
tail [or a $\Psi\propto (\Delta t)^{-l-2}$ tail in the case of static
initial setup].

There is yet another important outcome from the analysis at null infinity:
In the accompanying paper we apply a simple method to determine the late
time behavior of the scalar wave outside the Schwarzschild black
hole at {\it any constant radius}, down to the event horizon. This
is made possible only if one is provided with the form of the wave
at null infinity, which shall be calculated in Schwarzschild
using the same iterative scheme developed in the present paper.
Thus, even thought the iterative scheme seems, in general, not effective
at small distances, it still constitute, in our approach, a crucial step in
establishing a complete description of the wave behavior at late
time.

Generally speaking, we may point out the following
advantages of the new technique:
\begin{itemize}
\item{Most important, the new scheme is the first one directly extensible
to realistic rotating black holes, as shall be described in \cite{Ours}}
\item{The iterative scheme provides a convenient formal framework for
studying the effect of the non-trivial curvature-induced part
(``$\delta V$'') of the effective potential on the evolution of the fields.
This effect is isolated in our approach by explicitly splitting
the wave into its `trivial' (flat space) part $\Psi_{0}$, and a
curvature-induced part $\Psi-\Psi_{0}$, in which each term $\Psi_{N}$
arises from $N$ scattering events off the curvature potential $\delta V$. }
\item{We already discussed the fact that the iterative series can be
formally viewed as an expansion of the scalar field in the parameter
$M/u_{0}$. This feature provides a convenient formal way
to explore and quantify the dependence of the wave behavior on the
initial distance of the (compact) perturbation.
(an issue somewhat overlooked in previous works.)}
\end{itemize}

Apparently, the iterative scheme developed in this paper can be
directly applied to any asymptotically-flat spherically-symmetric
spacetime (possessing an effective potential barrier for the waves).
That would basically require only the modification of the curvature
potential
$\delta V$ to that characterizing the new spacetime under consideration.
The results of our analysis were consistent with the general
assumption that the small distance details of spacetime structure
do not affect the behavior of the wave at asymptotically late
time. (This was demonstrated for $\Psi_{1}$ and $\Psi_{2}$ by an
explicit calculation).
Thus it seems reasonable to assume that the characteristics of
this behavior would be completely determined by merely the large-$r$
asymptotic form of $\delta V$.

In the accompanying paper we use the iterative scheme to analyze
the behavior of scalar waves at null infinity in the 'complete'
Schwarzschild manifold. In this case the
details of the analysis become somewhat complicated by the presence of
the highly curved region near the horizon. Yet, we shall find that at
late time the wave behaves essentially the same as in the shell model,
in consistency with the above indications.

\section*{acknowledgments}

The author wishes to express his indebtedness to Professor A.\ Ori
for his guidance throughout the execution of this research and for
countless helpful discussions.

\appendix

\section{Uniqueness of the boundary condition problem} \label{appA}
The purpose of this appendix is to show that the initial value
problem for the scalar wave, with a boundary condition, as defined in Eqs.\
(\ref{eq4}), (\ref{eq6}) and (\ref{eq6a}), has a unique solution.
(The formal issue of uniqueness is not trivial because of the
divergence of the potential function at the center of symmetry
in the 1+1 representation.) It will also be verified that each of the
components $\Psi_{N}$ of our iterative expansion is uniquely
defined.

\subsection*{Uniqueness of $\Psi_{0}$}

First, it will be shown that the solution for
$\Psi_{0}$ (the ``Minkowski-like'' first component of the iterative
expansion), stated in Eq.\ (\ref{eq20}), is unique.
To that end we examine the most general solution to the wave
equation (\ref{eq17}), given by
\begin{equation} \label{eqA1}
\Psi_{0}=\sum_{n=0}^{l}A_{n}^{l}\frac{g_{0}^{(n)}(u)+
(-1)^{n}h_{0}^{(n)}(v)}{(v-u)^{l-n}},
\end{equation}
in which the coefficients $A_{n}^{l}$ are those given in
Eq.\ (\ref{eq21}), and where $g(u)$ and $h(v)$ are (yet) arbitrary
functions. (In flat spacetime, these two functions describe the outgoing
and ingoing components of the wave, respectively).
It will now be shown that although the functions $g(u)$ and $h(v)$
are not uniquely determined by the initial and boundary
conditions, the wave $\Psi_{0}$ {\em is} unique.

Consider first the evolution of the wave between retarded times
$u=u_{0}$ and $u=0$ (see figure \ref{fig2} for reference).
In this region the wave equation possesses no irregular
points, and thus existence and uniqueness of the solution $\Psi_{0}$
is guaranteed relying on standard theorems (see, for example,
\cite{Friedlander75}). Therefore, the solution stated
in Eq.\ (\ref{eq20a}) is unique at $u\leq 0$.

We thus focus on the region $u\geq 0$, where we have to show that
with the conditions $\Psi_{0}(u=0)=0$ and $\Psi_{0}(r=0)$ imposed
on the general solution $(\ref{eqA1})$, the trivial solution
$\Psi_{0}(u\geq 0)\equiv 0$ is unique. (Due to the linearity of the wave
equation, to verify uniqueness it is
enough to consider the case in which the wave vanishes on the
initial ray $u=0$\footnote{For, if there
exist two solutions at $u\geq 0$, both subject to the same initial
and boundary conditions, then their difference admits the same wave
equation, with zero initial and boundary conditions. Then, showing
that the difference is identically zero implies that the two
original solutions must coincide.}.
Thus the use of the condition $\Psi_{0}(u=0)=0$ in this context is
regardless of the fact that the wave
$\Psi_{0}$ actually vanishes along the outgoing ray $u=0$.)

Imposing the condition $\Psi_{0}(u=0)=0$ on the general solution
(\ref{eqA1})
leads to a linear inhomogeneous ordinary equation for the function
$h(v)$, analogous to Eq.\ (\ref{eq21b}). By analogy to Eq.\
(\ref{eq21a}), the general solution to this equation reads
\begin{eqnarray} \label{eqA2}
h(v)&=&\frac{1}{(l-1)!}\sum_{n=0}^{l} A_{n}^{l}g_{0}^{(n)}(0)\;v^{l+1}
\!\int_{v}^{\infty}\frac{(v-v')^{l-1}}{(v')^{2l-n+1}}dv' \nonumber\\
    & & + \sum_{n=l+1}^{2l} \frac{1}{n!}c_{n}v^{n},
\end{eqnarray}
where $c_{n}$ are $l$ arbitrary parameters. Upon calculating the
integral, we find that $h(v)$ is just a polynomial of order $2l$,
\begin{equation} \label{eqA3}
h(v)=\sum_{n=0}^{2l} \frac{1}{n!} c_{n} v^{n},
\end{equation}
with its first $l+1$ coefficients given by
\begin{equation} \label{eqA4}
c_{n}=-g_{0}^{(n)}(0) \mbox{\ \ \ \ \ for $0\leq n\leq l$}.
\end{equation}

Next, we require that $\Psi_{0}$ would vanish at $r_{*}=0$ (that
is for $u=v$). By Eq.\ (\ref{eqA1}) this requirement makes the
functional equality $g(u)=-h(u)$ compulsory. Note that this result
accommodates with Eqs.\ (\ref{eqA3},\ref{eqA4}),
allowing both conditions $\Psi_{0}(u=0)=0$
and $\Psi_{0}(r_{*}=0)=0$ to be satisfied simultaneously without
forcing the functions $g(u)$ and $h(v)$ to vanish identically.

We conclude that $\Psi_{0}$ has the form
\begin{equation} \label{eqA5}
\Psi_{0}=\sum_{n=0}^{l}A_{n}^{l}\frac{{\cal P}_{l}^{(n)}(u)-
(-1)^{n}{\cal P}_{l}^{(n)}(v)}{(v-u)^{l-n}},
\end{equation}
where ${\cal P}_{l}$ are polynomials of order $l$ [the $(l+1)$th
through $(2l)$th powers of the polynomials vanish independently in
the summation, in view of Eq.\ (\ref{eq23})]. Notice that formally,
$\Psi_{0}$ vanishes along the ray $u=0$ for any
polynomial ${\cal P}_{l}$ whatsoever, as the polynomial
coefficients $c_{n}$ are arbitrary in our construction.

The last remark is important, since it directly implies that
$\Psi_{0}$ vanishes {\em anywhere} at $u\geq 0$. For, given a
specific value of retarded time $\bar{u}>0$, we may transform to
the new coordinates $x\equiv u-\bar{u}$ and $y\equiv v-\bar{u}$,
in terms of which we have
\begin{eqnarray} \label{eqA6}
\Psi_{0}&=&\sum_{n=0}^{l}A_{n}^{l}\frac{{\cal P}_{l}^{(n)}(x+\bar{u})-
(-1)^{n}{\cal P}_{l}^{(n)}(y+\bar{u})}{(y-x)^{l-n}} \nonumber\\
&=& \sum_{n=0}^{l}A_{n}^{l}\frac{\bar{{\cal P}}_{l}^{(n)}(x)-
(-1)^{n}\bar{{\cal P}}_{l}^{(n)}(y)}{(y-x)^{l-n}},
\end{eqnarray}
where $\bar{{\cal P}}_{l}$ is still another polynomial of order
$l$. The last expression vanishes at $x=0$ (for any $v$) regardless of
the form of the polynomial, and thus we find that $\Psi_{0}$ vanishes
also along
the ray $u=\bar{u}$. Since the value of $\bar{u}$ is arbitrary, we
are led to the conclusion that $\Psi_{0}$ must vanish identically at
$u\geq 0$.
Therefore, the solution (\ref{eq20}) for $\Psi_{0}$ is unique.

\subsection*{Uniqueness of the functions $\Psi_{N}$}

To show that the functions $\Psi_{N}$, for $N\geq 1$, are unique,
it is enough to verify that the Green's function $G$ is unique.
Then, all function $\Psi_{N}$ are unique by construction.

To discuss the uniqueness of the Green's function, we refer to a
source point at $(u',v')$, as sketched in figure \ref{fig3}.
In region I (see the figure) existence and uniqueness are
guaranteed by standard theorems \cite{Friedlander75}.
In region II the Green's function admits the general form
(\ref{eqA1}), as do $\Psi_{0}$ (both functions obey the same
homogeneous field equation in this region).
In the expression corresponding to the Green's function we shall
denote the two functions of $u$ and $v$ by $g_{G}(u)$ and
$h_{G}(v)$ respectively.

The value of the Green's function along the outgoing ray $u=v'$,
dictated by the evolution in region I, is unity.
Hence, if the boundary condition $G(r_{*}=0)=0$ is to be
satisfied, then there has to be a discontinuity in the Green's
function at retarded time $u=v=v'$ (the left vertex
of the dark-colored rectangle in figure \ref{fig3}).

It can be shown that no solution exist for the Green's function,
which is continuous along $u=v'$ (with the single point $u=v=v'$ excluded)
and, at the same time, satisfies the boundary condition.
For, if we assume that $G(u\rightarrow (v')^{+})=1$, then we shall
have the function $h_{G}$ admitting the form
\begin{eqnarray} \label{eqA7}
h_{G}(v)=\sum_{n=0}^{2l} \frac{1}{n!} c_{n} (v-v')^{n}
\end{eqnarray}
[analogous to Eq.\ (\ref{eqA3})],
where this time the coefficients $c_{n}$ shall read
\begin{eqnarray} \label{eqA8}
c_{n}=\left\{\begin{array}{lc}
             -g_{G}^{(n)}(v')    &  ,0\leq n\leq l-1 \\
             -g_{G}^{(n)}(v')+1  &  ,n=l
            \end{array}.
      \right.
\end{eqnarray}
This cannot be accommodated with the requirement that
$h_{G}(u)=-g_{G}(u)$, necessary for the boundary condition to
hold.
Thus the Green's function cannot be continuous along $u=v'$.

Now, the Green's function can still obey the field
equation at $u=v'$, provided that the discontinuity
along this ray is constant in magnitude for all $v>v'$.
This can be verified by writing
$G=G\times \theta(u-v')+G\times \theta(v'-u)$,
and substituting into the equation
for the Green's function. One thereby finds that for this
equation to be satisfied at $u=v'$, we must have
$G,_{v}(u\rightarrow (v')^{+})=0$, and thus
$G(u\rightarrow (v')^{+})=k$, where $k$ is constant.
By analogy to Eq.\ (\ref{eqA8}) we find that in this case
$h_{G}^{(l)}(v')=-g_{G}^{(l)}(v')+k$, which implies that the boundary
condition is violated, unless $k=0$.
Hence, the Green's function must ``jump'' to zero right after
retarded time $u=v'$.

Then, by the discussion regarding $\Psi_{0}$,
we conclude that the only solution to the Green's function in
region II, subject to the boundary condition at the origin and
continuous everywhere except on the ray $u=v'$, is the trivial
solution $G(u>v')\equiv 0$.

\subsection*{Uniqueness of $\Psi$}

Now, consider the initial value (and boundary condition) problem
for the ``complete'' scalar wave $\Psi$ in the shell model.
To verify that a solution is unique, again we analyze the situation of
zero initial data (on $u=u_{0}$ and $v=0$), showing that in this
case the null solution ($\Psi\equiv 0$) is unique.

At $u\leq 0$, existence and uniqueness of $\Psi$ is assured, relying
on standard theorems. Thus the trivial solution $\Psi(u\leq 0)\equiv 0$
is unique. Particularly, we find that $\Psi$ vanishes along the
ray $u=0$.

\begin{figure}[htb]
\input{epsf}
\centerline{\epsfysize 6cm \epsfbox{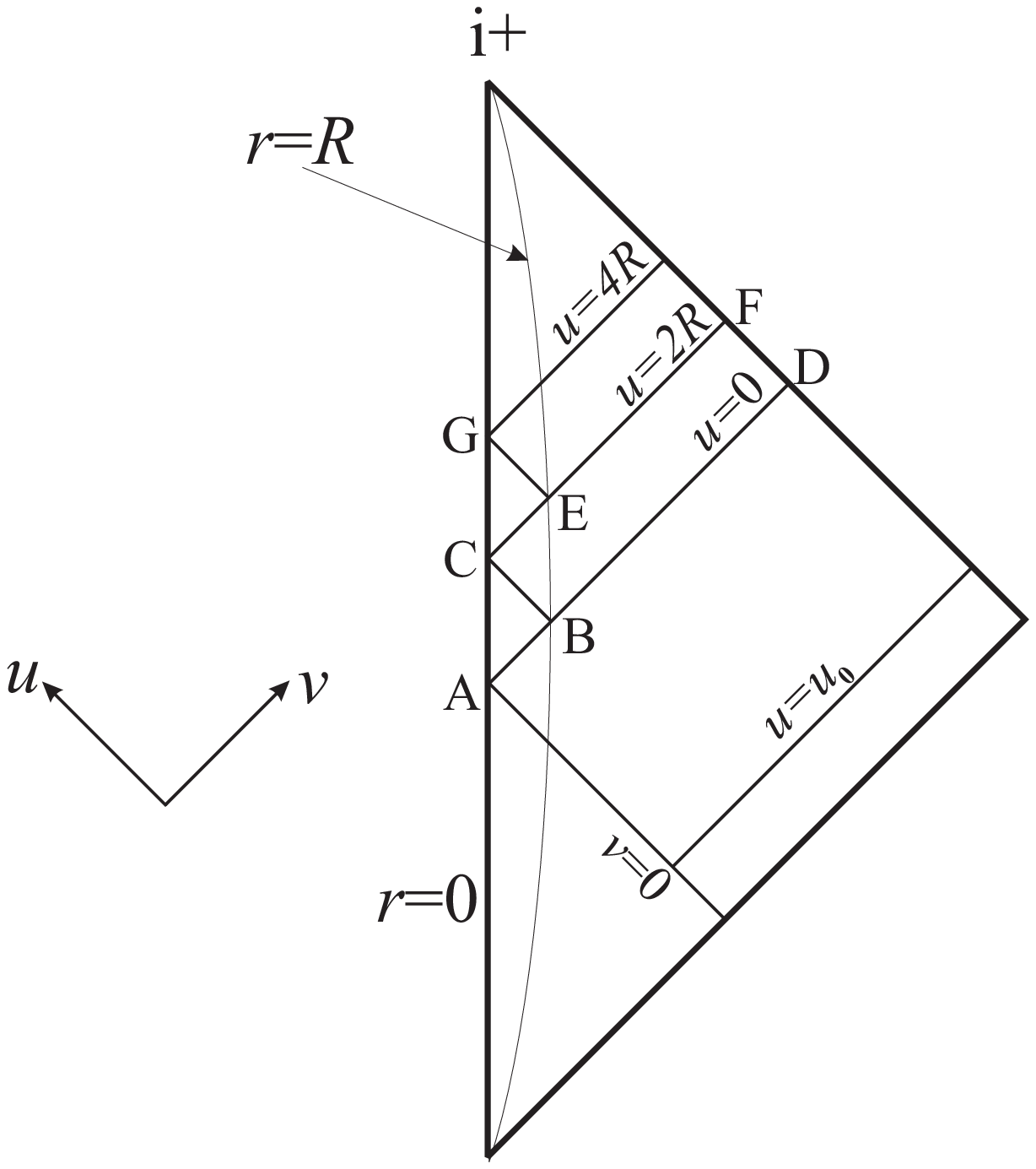}}
\caption{\protect\footnotesize
Uniqueness of the solution for $\Psi$ (see explanation in the
text).}
\label{fig12}
\end{figure}

Now, consider the region $u\geq 0$, $v\leq 2R$,
represented in the diagram of figure \ref{fig12} by the triangle
$ABC$. In this portion of spacetime (confined to the flat interior
of the shell) $\Psi$
obeys the same equation as $\Psi_{0}$, and is subject to the
boundary condition at $r_{*}=0$ and to zero initial condition
along the null segment $AB$ (see the figure). By the above
discussion (regarding $\Psi_{0}$) we deduce that $\Psi$ must
vanish in the portion $ABC$. In particular, we find that $\Psi$
is zero along the null segment $BC$. This, in turn,
together with the fact that the wave vanishes along the ray $BD$
(see figure), implies (again by standard theorems) that $\Psi$ vanishes
anywhere inside the region represented in figure \ref{fig12} by
the rectangle $BDFC$. Thereby we have shown that the wave vanishes
anywhere between retarded times $u=0$ and $u=2R$.

We can now proceed, referring next to the triangle $CEG$ (see
figure), to show that $\Psi$ vanishes also
between retarded times $u=2R$ and $u=4R$. The same procedure may
be applied an arbitrary number of times, showing that the wave $\Psi$
must vanish anywhere. Since the wave equation is linear, we have
thereby shown that the solution for $\Psi$ (with arbitrary initial
data and zero boundary condition at the center of symmetry) is unique.

\section{Derivation of $\lowercase{g}_{1}^{\infty}$ for
         $\lowercase{l}\geq 1$} \label{appB}

In this appendix we give a detailed derivation of $g_{1}^{\infty}(u)$
[Eq.\ (\ref{eq46})] for the case $l\geq 1$.
The starting point to our calculation is Eq.\ (\ref{eq43}).
It has been shown that the $r_{*}^{-3}$ potential [namely the term
proportional to $a$ in Eq.\ (\ref{eq41})] gives no contribution to
$g_{1}^{\infty}$.
We therefore set $a=0$, then integrate Eq.\ (\ref{eq43}) by parts once
with respect to $v'$. Using Eq.\ (\ref{eq32a}), this yields
\begin{eqnarray} \label{eqB1}
g_{1}^{\infty}(u) & = &
-\frac{b}{l!}\sum_{n=0}^{l} B_{n}^{l}
\int_{u_{0}}^{u}du'g_{0}^{(n)}(u')(u-u')^{n-2}           \nonumber\\
& & \times
\left[\ln (\tilde{v}'-\tilde{u}')+\sum_{m=0}^{l}\frac{l!}{2l-n-m+2}
\right],
\end{eqnarray}
with the coefficients $B_{n}^{l}$ given in Eq.\ (\ref{eq45}).

To proceed, consider first the logarithmic part of Eq.\ (\ref{eqB1}).
We integrate by parts each of the terms in this part $n$ successive times
with respect to $u'$. As explained previously, compactness of
$g_{0}(u)$ assures that the resulting surface terms would all vanish.
One is thus left with
\begin{equation} \label{eqB2}
-\frac{b}{l!}\sum_{n=0}^{l} B_{n}^{l}(-1)^{n}
\int_{u_{0}}^{0}du'g_{0}(u')\frac{\partial^{n}}{\partial u'^{n}}
\left[\frac{\ln(\tilde{v}'-\tilde{u}')}{(u-u')^{2-n}}\right].
\end{equation}

Summing first the $n=0$ and $n=1$ terms yields
\begin{equation} \label{eqB3}
-\frac{b}{2(2l+1)}\int_{u_{0}}^{0}\frac{g_{0}(u')}{(u-u')^{2}}du',
\end{equation}
in which the vanishing of the logarithmic dependence arises from the
equality of $B_{0}^{l}$ and $B_{1}^{l}$, which was also responsible for the
vanishing of the contribution from scattering off the ``$r_{*}^{-3}$''
potential.

The remaining terms in the sum in Eq.\ (\ref{eqB2}) take the form
\begin{eqnarray} \label{eqB4}
b\sum_{n=2}^{l}\sum_{j=0}^{n-2} B_{n}^{l}
\frac {n!(n-2)!(-1)^{n+j}}{j! (n-j-2)! (n-j)}
\int_{u_{0}}^{0}\frac{g_{0}(u')}{(u-u')^2}du'               \nonumber\\
=b\frac{l-1}{4l(l+1)}\int_{u_{0}}^{0}\frac{g_{0}(u')}{(u-u')^{2}}du'
\end{eqnarray}
(where the expression given for the sums over $j$ and $n$ is not too
difficult to verify, using simple combinatorial manipulations).

The sum of two expressions given in (\ref{eqB3}) and (\ref{eqB4})
corresponds to the logarithmic part of Eq.\ (\ref{eqB1}).
We still have to consider the contribution of the non-logarithmic terms.
Integrating these terms by parts $n$ times with respect to $u'$
(with the resulting surface terms again vanishing), we find that only the
$n=0$ and $n=1$ terms survive, with their sum given by
\begin{eqnarray} \label{eqB5}
-\frac{b}{2(2l+1)}\sum_{m=0}^{l}\left[\frac{1}{2l-m+3}-\frac{1}{2l-m+2}
\right]                                        \nonumber\\
\times \int_{u_{0}}^{0}\frac{g_{0}(u')}{(u-u')^{2}} du',
\end{eqnarray}
which is
\begin{equation} \label{eqB6}
\frac{b}{4(l+1)(2l+1)}\int_{u_{0}}^{0}\frac{g_{0}(u')}{(u-u')^{2}} du'.
\end{equation}

We finally obtain $g_{1}^{\infty}(u)$ by collecting the expressions
(\ref{eqB3}), (\ref{eqB4}) and (\ref{eqB6}), and setting $b=8Ml(l+1)$
[by Eq.\ (\ref{eq42})]. This produces Eq.\ (\ref{eq46})

\section{\mbox{}\\Contribution to $\lowercase{g}_{2}^{\infty}$ due to
$\Delta_{1}(\lowercase{u}',\lowercase{v}')$} \label{appC}

The purpose of this appendix is to prove Eq.\ (\ref{eq64}). This equation
sets an upper bound on the contribution to $g_{2}^{\infty}$ due to the
$v$-dependent part of $g_{1}$, namely $\Delta_{1}(u,v)$ (defined in
Eq.\ (\ref{eq63})).

We first show that the following upper bound is applicable to
$\Delta_{1}(u,v)$ at large $r$:
\begin{equation} \label{eqC1}
\left|\Delta_{1}(u,v)\right|\leq CMr_{*}^{-2}\ln \tilde{r}_{*},
\end{equation}
in which $C$ is a positive constant.
To verify the validity of this inequality, examine the explicit form of
$\Delta_{1}$, derived directly from Eq.\ (\ref{eq56}),
\begin{eqnarray} \label{eqC2}
\Delta_{1}(u,v)=\sum_{k=0}^{l} \int_{u_{0}}^{0}\!\!\!du'
\int_{v}^{\infty}\!\!\!dv'\frac{(v'-u)^{l} (u-u')^{l-k}}{(v'-u')^{2l-k+3}}
                                                               \nonumber\\
\times \left[a_{k}^{l}+b_{k}^{l}\ln(\tilde{v}'-\tilde{u}')\right] g_{0}(u').
\end{eqnarray}

Now, since $(v'-u)\leq (v'-u')$ and $(u-u')\leq(v'-u')$, we find that there
exist positive numbers $C'$, $C''$, $C'''$ and $C$, such that
\begin{eqnarray} \label{eqC3}
\lefteqn{\left|\Delta_{1}(u,v)\right|\leq}                        \nonumber\\
& & C'M\int_{u_{0}}^{0}du'\int_{v}^{\infty}dv'
\frac{\ln (\tilde{v}'-\tilde{u}')}{(v'-u')^{3}}\left|g_{0}^{\infty}(u')
\right|\leq\nonumber\\
& & C''M\int_{u_{0}}^{0}du'
\frac{\ln (\tilde{v}-\tilde{u}')}{(v-u')^{2}}\left|g_{0}^{\infty}(u')
\right|\leq  \nonumber\\
& & C'''M \frac{\ln (\tilde{v}-\tilde{u})}{(v-u)^{2}}
\int_{u_{0}}^{0}du'\left|g_{0}^{\infty}(u')
\right|\leq    \nonumber\\
& & \hspace{40mm} CMr_{*}^{-2}\ln \tilde{r}_{*},
\end{eqnarray}
which confirms Eq.\ (\ref{eqC1}).

We are now in position to evaluate the
contribution of $\Delta_{1}$ to $g_{2}^{\infty}(u\gg M)$.
By Eqs.\ (\ref{eq56}) and (\ref{eqC1}) we deduce that this contribution
(in absolute value) is bounded from above by
\begin{equation} \label{eqC4}
M^{2}\sum_{k=0}^{l}C\int_{u_{0}}^{0}\!du'\int_{u}^{\infty}\!dv'
\frac{(v'-u)^{l}(u-u')^{l-k}}{(v'-u')^{2l-k+5}}
\ln^{2}(\tilde{v}'-\tilde{u}').
\end{equation}
Since $(v'-u)\leq (v'-u')$ and $(u-u')\leq (v'-u')$, we find, in turn,
that the last expression is bounded from above by
\begin{eqnarray} \label{eqC5}
C'M^{2}\int_{u_{0}}^{0}\!\!\!du'\int_{u}^{\infty}\!\!\!dv'
\frac{\ln^{2}(\tilde{v}'-\tilde{u}')}{(v'-u')^{5}} & \leq   & \nonumber \\
C''\int_{u_{0}}^{0}\!\!\!du' \frac{\ln^{2}(u-u')}{(u-u')^{4}} &\leq &
CM^{2}u^{-3}\ln^{2} \tilde{u},
\end{eqnarray}
for $u\gg M$. [Here, $C'$, $C''$ and $C$ are some positive numbers, other
then in Eq.\ (\ref{eqC3})].

The inequality (\ref{eq64}) is thereby proved.



\end{document}